\documentclass{article}

\usepackage{PRIMEarxiv}

\usepackage[utf8]{inputenc} % allow utf-8 input
\usepackage[T1]{fontenc}    % use 8-bit T1 fonts
\usepackage{hyperref}       % hyperlinks
\usepackage{url}            % simple URL typesetting
\usepackage{booktabs}       % professional-quality tables
\usepackage{amsfonts}       % blackboard math symbols
\usepackage{nicefrac}       % compact symbols for 1/2, etc.
\usepackage{microtype}      % microtypography
\usepackage{lipsum}
\usepackage{fancyhdr}       % header
\usepackage{graphicx}       % graphics
\graphicspath{{media/}}     % organize your images and other figures under media/ folder

\title{Systematic Evaluation and User Study of Privacy of Default Apps in Apple's Mobile Ecosystem}
\author{
  Amel Bourdoucen\\
  Aalto University \\
   \And
  Janne Lindqvist \\
  Aalto University \\}
\begin{document}
\maketitle

\begin{abstract}
Users need to configure default apps when they first start using their devices.  The privacy configurations of the default apps do not always match what users think they have initially enabled. We first systematically evaluated the privacy configurations of default apps. We discovered serious issues with the documentation of the default apps. Based on these findings, we explored users' experiences with an interview study (N=15). Our findings from both studies show that: the instructions of setting privacy configurations of default apps are vague and lack required steps; users were unable to disable default apps from accessing their personal information; users assumed they were being tracked by some default apps; default apps may cause tensions in family relationships because of information sharing. Our results illuminate on the privacy and security implications of configuring the privacy of default apps and how users perceive and understand the mobile ecosystem.
\end{abstract}

% keywords can be removed

%\keywords{Privacy, Mobile Devices, Apps, Ecosystems.}

%\section*{Acknowledgments}
%-------------------------------------------------------------------------------
\section{Introduction}
%-------------------------------------------------------------------------------
%================================================================================
%====================== INTRODUCTION =================================================
%================================================================================

Users are not in control of their privacy preferences in complex mobile device and cloud ecosystems.
The complexity of these systems enable helpful features, but with the cost of privacy. Often, many of these features are enabled by default when a user first starts using the device(s). When users purchase new devices, they are often presented with many features to enable or disable. Unfortunately, our work shows that the privacy implications of these features are not understood and the settings are non-trivial to configure correctly.

Studies have discovered several challenges that users face in understanding how their data is handled when using mobile apps \cite{notechnicalunderstandingrequired,improvingpermissions:palmerino,privacyprofiles:liu,Van2017BetterTheDevilYouKnow,forgivness:thompson}. Users are often presented with many permissions requesting approval of access to their personal data. Moreover, prior work suggests that users do not always fully understand the implications of enabling privacy configurations \cite{smartphoneprivacy:2022:Frik,survey:breitinger:2019}. This confusion is often caused by the architecture of privacy configurations \cite{priceforthat:serge:2013}. Users are left confused \cite{joshuaexplanationforpermission2014}, surprised \cite{Balbeko2013LittleBrothersWatchingYou,Leakinessandcreepiness:irina}, anxious \cite{Ifeelinvadedannoyed:2022:usenix} and sometimes frustrated \cite{Leakinessandcreepiness:irina} when learning what actually happens to their data. %% jl: how our work is different ending sentence

In this paper, we study default apps in Apple's iOS and macOS\footnote{
This work is an independent publication and has not been authorized, sponsored, or otherwise approved by Apple Inc.}. MacOS and iOS include setup wizards that guide users to selecting preferences for these default apps when a new device is started. As a concrete example, in the Setup Wizard, one of the features that users can setup is \textit{Siri}. Users are offered the choice to either \textit{Continue} (to set up Siri) or \textit{Set Up Later in Settings}. Users can select \textit{Set Up Later in Settings}, which implies that Siri is disabled until users set it up later. 

This is despite selecting the option to \textit{Set Up Later in Settings.} To systematically study these issues, we focused on the following research questions:

\begin{itemize}
    \item[] \textbf{RQ1:} What privacy configurations are available to control default apps?
    \item[] \textbf{RQ2:} How can users control default apps?
    \item[] \textbf{RQ3:} How do users understand privacy configurations and their privacy and security implications?
    \item[] \textbf{RQ4:} How does setting up default features impact privacy of users?
\end{itemize}

To answer these research questions, we conducted two studies; in the first study we selected eight default apps and features on iOS and macOS: Safari (Web browsing), Family Sharing (Shared Access), Siri (Virtual Assistant), iMessage (Messaging), Facetime (Video Calls), Location Services and TouchID (Fingerprint). We analyzed Apple's official documentation and found them to seriously lack required details for configuring privacy. Due to this, we conducted a comprehensive system evaluation to understand the privacy configurations for these default apps. We mapped the routes to disable features of these apps. Based on this first study, we conducted a follow-up user study to investigate users' understanding of privacy configurations. We compared the users answers to our findings from Study 1.
%We discovered users' expectations about data handling processes and compared them to our findings of what actually occurs to their data. 

We make the following major contributions:

\begin{enumerate}
    \item In our first study, we thoroughly collected and investigated the privacy configurations of eight default apps of the Apple mobile ecosystem. We present what user data is collected by each default app as well as paths to disable different privacy configurations (presented in Section \ref{study1-sec}). We observed several serious issues associated with configuring default apps.
    %\item In the second study, we discovered users' understanding of using default apps, for example, what data is shared when using an app (presented in Section \ref{study2-sec}).
    \item In the second study, we conducted a user study to explore users' perceptions of using default apps and configuring their privacy and security settings. Our findings show that users had limited understanding of what data was being collected and how it was shared (presented in Section \ref{discussion-sec}).
    \item Based on the two studies, we made the following major findings: participants were surprised to know some privacy configurations were turned on by default; participants were confused about what happened to their information; some participants assumed they were being tracked by some apps; tracking can introduce trust issues in families and relationships; participants were aware that disabling data sharing does not guarantee data is not shared anymore;  participants wanted to know what happens to their information; participants did not know how to disable data sharing on default apps.
\end{enumerate}

%-----------------------------------
\section{Related Work}
\label{Relatedwork}

%================================================================================
%====================== LITERATURE =================================================
%================================================================================

We first discuss the most closely related prior work and proceed to discuss studies that explored users general understanding of privacy in apps. We then describe challenges that users face when setting privacy configurations to gain 
control over their personal data. Most closely related to our work are the following three studies:

%Most closely related to our work are three studies that explored privacy configurations in pre-installed apps \cite{smartphoneprivacy:2022:Frik,Gamba:2020:preinstalledapps,Ramokapane2019}.

First, a study \cite{smartphoneprivacy:2022:Frik} explored the relationship between socio-economic factors and users’ choices of security and privacy settings. The study surveyed users about the security and privacy settings of their mobile device’s operating system (Android and iOS) and pre-installed browsers (Safari and Chrome). These settings included passcode, face unlock, automatic updates, and password re-use. Participants were also asked about privacy and security risks the are concerned of. The study found differences between socio-demographic groups on using security and privacy settings. For example, older adults were found to worry less about online risks. The study also found that many users were not aware of the function of settings used in the study but were willing to change them in the future. %The study relied on participants’ self-reported answers to analyze their knowledge of privacy and security settings on their mobile devices. 

Second, another study~\cite{Gamba:2020:preinstalledapps} explored pre-installed apps on the Android platform for their app packages, certificates, and third-party libraries. This work revealed various actors involved in the development of pre-installed apps. Potentially harmful behavior was detected from pre-installed apps related to personal data collection. The study validated which app collected which set of personal information by collecting firmware and traffic information from users. 
%Users’ activities on their mobile devices are monitored by parties  that users’ may not be aware of. 

Third, another study~\cite{Ramokapane2019} explored users' awareness of some features on both Android and iOS. These features were 1) location, 2) ads tracking and 3) usage and diagnostics (this is called Analytics Data Sharing in iOS). The study asked participants to conduct four tasks on either a iOS or Android device based on the participant's experience and preference. These tasks were: Disable location services, Restrict App from using a default feature (iOS: Camera, Android: disable Google app having access to Calendar and Location services), Disable/Limit ad tracking, Restrict Usage and Diagnostic Report (Android only) or Disable analytical data sharing (iOS only). The participants were given devices owned by the authors that were reset so that these tasks to be completed. The authors enabled the settings of the features and asked users to locate and disable them. The participants performed a cognitive walk trough and think aloud protocols while performing these tasks. The findings showed that participants were not able to easily locate the settings of the features used in the study. Participants attributed these challenges to hidden controls and complex app requests. To overcome these challenges, participants used quick fixes or coping strategies such as skipping privacy configuration requests or searching the internet for quick solutions.

%In this work, we explore the privacy and security settings of eight default apps including browsing app Safari on Apple devices. To combat possible issues with self-reported answers, we asked participants to find and configure settings of default apps using their devices. We highlight several challenges to locate and disable features of default apps.

In contrast to the above three prior work, we evaluated the privacy configurations available in the Apple iOS and macOS systems and specifically probed into users' understanding of privacy in the Apple ecosystem. Towards this end, we systematically tested setting up both iOS and macOS devices and documented all the steps required to disable settings. We used this information as the basis of our qualitative interviews towards the end of probing to users' understandings and experiences with these privacy configurations and devices. We asked our participants to use their own devices rather than test devices provided by us. We wanted our participants to perform the tasks with i) familiar devices and ii) devices that contain their own privacy configurations that they have setup. Our results reflected users' actual experiences with privacy settings they have configured on their own rather than if we pre-configured it for them. Our work contributes by revealing the difficulties on properly configure privacy for features that are enabled by default, and the surprises and tensions that users experiences because of these settings.

%In contrast, our study \textcolor{red}{explores users awareness of data sharing of eight default apps part of Apple's Cloud system. We asked participants to use their own devices. Participants using their own devices and iCloud accounts rather than test devices provided by us meant that participants were more familiar with their own devices. Furthermore, participants were able to reflect on privacy configurations that they have setup themselves at a point in time. We highlight users' difficulties to disable any of the default apps.  As well as, challenges associated with using default apps in a family setting.}

Next, we will discuss a body of research that explored privacy of users when using mobile apps and configuring privacy settings.

%%============================================================
\paragraph{User Privacy in Mobile Apps}
%% 1) what fields have been explored in mobile app privacy
Prior work has largely explored app permissions in mobile devices. Studies have been motivated to explore users configuring permissions due to the reported difficulties when doing so. For several years, researchers have focused on permissions of apps in an attempt to explore ways to improve users' understanding and expectations when setting privacy configurations of apps. The focus on app permissions was motivated through the many studies that demonstrated users' difficulties in understanding privacy configurations of mobile apps \cite{Balbeko2013LittleBrothersWatchingYou,Ramokapane2019, joshuaexplanationforpermission2014,notechnicalunderstandingrequired}. 

A study on 308 Android users revealed that only 17\% of users were attentive to the permissions that were prompted during app installations, therefore indicated that permission warnings are not sufficient to make informed security decisions \cite{feltandroidpermissions2011}. A recent study in 2021 on 4,636 Android users has also confirmed that information provided by the current system is not enough for users to make informed decisions on their privacy \cite{Shen2021}. Other studies also showed that often users either ignore or accept permissions without reading the details properly \cite{felt201299problems,Felt2012androidpermissions,Ramokapane2019}. 
%\newline

Researchers highlighted factors that influence the misunderstandings that users have of privacy configurations. Several factors make it harder for users to know what happens to this personal information when agreeing to permissions or configurations; for example, unclear Privacy Policies, and lack of transparency about data practices. Earlier studies have investigated ways to improve Privacy Policies for better delivery for users \cite{Kelley2009Nutritionlabel}. However, recent studies have reported that the unclear nature of Privacy Policies of apps still contribute to the difficulty users have to grasp what happens to their personal data \cite{Alohaly2016quantificationprivacypolicies,Coen2016paradox,Kelley2012conundrumofpermissions}. As a result of the unclear nature of Privacy Policies, users rarely follow Privacy Policy links to read what part of their information is disclosed \cite{Coen2016paradox}. 
%\newline
Another factor that has been suggested to contribute to the difficulty in understanding what happens to personal data in apps is transparency \cite{notechnicalunderstandingrequired,Van2017BetterTheDevilYouKnow}. A study found that providing more transparency to users about what occurs to their personal data can make users more confident in their app use \cite{Van2017BetterTheDevilYouKnow}.
To help users better understand how their personal data is handled, recent literature work explored several solutions to help make more informed decisions \cite{notechnicalunderstandingrequired,privacyprofiles:liu,improvingpermissions:palmerino,forgivness:thompson,Van2017BetterTheDevilYouKnow}. For instance, a study deployed machine learning to offer a prospect of mitigating the burden of increased privacy decisions \cite{smullen:2020}. Another study proposed a prototype that adjusted privileges given to apps on iOS as well as the ability to replace real data with mock data \cite{rethinkingapppermissions}. Another study analysed settings of 4.8 million smartphone users and demonstrated a number of profiles that aim to simplify the decisions mobile users have to make about their privacy \cite{privacyprofiles:liu}.

%However, users awareness of default app permissions does not necessarily imply that users can easily locate or are aware of the implications of these privacy settings \cite{Ramokapane2019}. In the same study, it is also observed in the case of default apps sharing with the ecosystem, where users are not necessarily aware of which apps share what information. 

%%============================================================

%%============================================================
\paragraph{Impacts of setting privacy configurations}
%%============================================================

A second line of research focused on understanding user's concerns when it comes to setting privacy preferences \cite{Balbeko2013LittleBrothersWatchingYou,Ramokapane2019,Shen2021,Wijesekera2018PrivacyDecisionsPredicition}. Research has found that users often have misconceptions about the data sharing that occurs on smartphone apps \cite{Balbeko2013LittleBrothersWatchingYou,smartphoneprivacy:2022:Frik,Ramokapane2019, joshuaexplanationforpermission2014}. Misconceptions about the handling of personal data can create challenges to users. Studies suggested that users may feel uncomfortable or confused when learning about what occurs to their data \cite{joshuaexplanationforpermission2014}. For example, studies suggested that users are often surprised when asked to share their personal data collected by apps \cite{Balbeko2013LittleBrothersWatchingYou,Leakinessandcreepiness:irina} for example, users understood that data was used for purposes such as marketing but were surprised by the scope of data sharing, frequency and destination \cite{Balbeko2013LittleBrothersWatchingYou}. 

Users can also experience other emotions such as confusion about certain personal data that is requested, sometimes dismay or outrage \cite{Leakinessandcreepiness:irina}. Studies that focused on tracking of users by apps \cite{Ifeelinvadedannoyed:2022:usenix,smartusefulcreepy:2012:blase,McDonald2010BeliefsAB:2010}, suggest that users can feel negatively under the perception of being tracked. These feelings can include: anger, distrust and anxiety. Often users would feel accepting of the fact that they are tracked, under certain conditions \cite{Ifeelinvadedannoyed:2022:usenix}. The reactions users had upon learning about how their data is handled can be denoted to the insufficient information on mobile apps to help users make informed decisions \cite{Ramokapane2019}. Insufficient information provided about data handling may also lead users to think that these permissions are required for apps to run \cite{smullen:2020} which influences users to accept them.

\paragraph{Summary}
Prior work has focused extensively on privacy settings of mobile devices. An example of this is users' attitudes towards permissions of different apps on mobile devices \cite{felt201299problems,Felt2012androidpermissions,Ramokapane2019}, tracking of apps \cite{Ifeelinvadedannoyed:2022:usenix,McDonald2010BeliefsAB:2010,smartusefulcreepy:2012:blase} and privacy configurations of apps \cite{Balbeko2013LittleBrothersWatchingYou,smartphoneprivacy:2022:Frik,Ramokapane2019,joshuaexplanationforpermission2014}. Although prior work focused extensively on users' attitudes when setting \textit{permissions of apps on mobile devices}, little research (presented above) has focused on users' experiences when setting up privacy configurations of default apps.

%\cite{Balbeko2013LittleBrothersWatchingYou,smartphoneprivacy:2022:Frik,Ramokapane2019, joshuaexplanationforpermission2014,notechnicalunderstandingrequired}. 

%-------------------------------------------------------------------------------

%%====================Study 1 ============================
\section{Study I: Mobile Ecosystem Evaluation}
\label{background}

%================================================================================
%====================== STUDY 1 =================================================
%================================================================================

\label{study1-sec}
\begin{table*}[ht]
  \centering
 \caption{Privacy configurations of default apps. These privacy configurations may lead to personal data of users being transferred outside the device as shown in the last column. For the full table, refer to Appendix \ref{appendixC}}
    \begin{tabular}{llll}
    \toprule
    \textbf{Default App} & \textbf{N Steps} & \textbf{Privacy Configurations} & \textbf{May transfer to Cloud or Vendor's Servers} \\
    \midrule
    Safari & N>12  & IP Address & Yes \\
          &       & Private Browsing & No \\
          &       & Web Page Translation & Translation locally, other data may leave device \\
          &       & iCloud Syncing & Yes \\
          &       & Preload Top Hit in Safari & Information not provided by vendor \\
          &       & Sending Information to Apple & Yes \\
          &       & History and Website Data & Yes \\
    \midrule
    Siri  & N>9   & Ask Siri & Yes \\
          &       & Integrated apps & Yes \\
          &       & Siri and Dictation & Yes \\
          &       & Siri Personalisation & Yes \\
          &       & iCloud Syncing & Yes \\
          &       & Location Services & Yes \\
          &       & Request History & Yes \\
 
    \midrule
    Facetime & N>7   & Enable Facetime & Actual calls No, Otherwise Yes \\
          &       & Caller ID & Yes \\
          &       & SharePlay & Information not provided by vendor \\
          &       & Speaking & Information not provided by vendor \\
          &       & FaceTime Live Photos & Information not provided by vendor \\
          &       & Blocked Contacts & Information not provided by vendor \\

    \bottomrule
    \end{tabular}%
  \label{table-ppr:privacyconfig}%
\end{table*}%

%%%%%%%%%%%%%%%%%%%%%%%%%%%%%%%%%%%%
%%% table of personal data that is collected from users

\begin{table*}[ht]
  \centering
  \caption{Users' personal data collected from default apps. Users are not able disable the collection of some of the data below for an app to function \cite{Applesprivacypolicy}.}
    \begin{tabular}{p{7em}p{33.665em}}
    \toprule
    \textbf{Default app} & \textbf{Users' personal data collected (not limited to list below)} \\
    \midrule
    Safari & IP address, sites you visit; {open tabs, tab groups, AutoFill information, Bookmarks, Reading List and History}, attribution reports, payment method information.  \\
    \midrule
    Siri  & Contact names, nicknames and relationships, Music and Podcasts, Names of your and your Family Sharing members' devices, Accessories, Homes, Scenes and Members of Shared Home in Home app, Labels for Items (e.g., people's names in Photos and Alarms), Name of apps and shortcuts. \\
    \midrule
    iMessage & Articles, TV shows, Music and Photos. \\
    \midrule
    Facetime & Facetime Calls (e.g., who was invited to call, device network configurations), Apps using Facetime, Phone numbers, email addresses associated with account. \\
    \midrule
    Family Sharing & Apple Watch serial number, cellular hardware identifiers, family member's {health, location and contact data}, view logs and screenshots from Apple Watch. \\
    \midrule
    Touch ID & 360-degree orientation fingerprint data, passcode. \\
    \midrule
    Location & Location data, Location Search Query, Geo-tagged locations of nearby WiFi hotspots, GPS data, travel speed, barometric pressure, places you recently been, IP. \\
    \midrule
    Find My & Participation in Find My network, device location, information about device, information about account. \\
    \bottomrule
    \end{tabular}%
  \label{tab-ppr:datacollected}%
\end{table*}%

%%%%%%%%%%%%%%%%%%%%%%%%%%%%%%%%%%%%
%% Review: motivation for using Apple's ecosystem here

In our first study, we analyze the main eight features of the Apple iCloud mobile ecosystem. Today, we have few mobile ecosystems that are popular, for example, Apple's \cite{applecosystem}, Google's \cite{googleecosystem} and Huawei's \cite{huaweiecosystem}. As the target of our work, we choose Apple's ecosystem for the following reasons: (1) Apple's devices are popularly purchased worldwide. In 2021, Apple reported a 65.6 billion USD revenue in iPhones only in the first quarter of the same year \cite{buisnessinsideronline2021}. (2) Apple's mobile ecosystem is uniquely cohesive and its integrated model provides a quality experience for users, alongside a stated emphasis on privacy and security. 
%\ref{mobilecosystemsstudy}.

\subsection{Mobile Ecosystem Structure}
\label{definitionofecosystem}

A mobile ecosystem consists of a set of units (devices) interacting with each other through exchange of information, resources and artifacts \cite{campbellecosystems2010}. For instance, Apple's mobile ecosystem consists of: iCloud (Apple's Cloud system) which brings together devices such as: iPad, iMac, Macbook, and iPhone. Default apps are central to users' participation in the ecosystem. The apps manufactured by Apple are referred to in this work by \textit{Default apps}. For instance, a popular app \textit{Safari} is used as a browser for Apple devices. Information stored in \textit{Safari} such as Bookmarks are exchanged between devices that are connected to the same iCloud account.

\subsection{Method of Study I}
\label{Method-study1}
We evaluated the system for the following parameters: defining mobile ecosystems' structure, privacy configurations of default apps, number of privacy configurations to disable within an app and types of personal data collected by these default apps and whether the personal data is transferred outside the device. We then analysed in depth Apple's eight default apps: Safari, Family Sharing, Find My, iMessage, Facetime, Siri, Location Services and TouchID. These apps are linked to simple configuration options presented to the user when the user first time starts using macOS or iOS device.

We had the following three major tasks in the analysis:
\begin{enumerate}
    \item \textit{Analysis of Official Sources:}  We first read the official sources provided by Apple \cite{Applesprivacypolicy}.
    %The sources provided by Apple come with various challenges such as being extremely lengthy and not containing enough description on privacy configurations. Moreover, these sources included very general description of apps and limited and scattered guidelines to privacy controls of different apps. 
    There were several challenges to perform this analysis; \textbf{a.} \textit{Closed Ecosystem:} Apple's ecosystem is \textit{closed}; meaning that some of the specifications about the processing of personal data by default apps are not disclosed. Examples of non-disclosure includes ambiguous phrasing such as  ``subsets of data stored'' \cite{Applesprivacypolicy} without indicating what the subset of data includes, how is it processed and for how long it is retained. \textbf{b.} \textit{Scattered information:} When reading Apple's Privacy Policy for the steps to disable a feature, we discovered that the controls of some apps are described under other apps. For example, \textit{Siri} has a specific section where its controls are described. However, \textit{Siri} can also be found under \textit{Safari Search}'s section as well as \textit{Dictation}. In summary, the public Privacy Policies are long and do not contain easily accessible information how to precisely control privacy configurations.
    %The scattering of this information can be confusing to the already long privacy policy. Regardless, it was important to highlight in this work what was readily available to users and what was not. The results of the study backed up this work to provide a complete understanding of the gap between users' knowledge of the data handling processes in the ecosystem and what is provided to users regarding data handling practices of the vendor.
    \item \textit{System Navigation:} To be able to present the steps required to setup a device, we captured the setup and usage processes on sample iOS and macOS devices. We first did factory resets to our test devices before following steps of the setup wizard in all possible combination of scenarios and sequence of steps. We repeated the latter process every time we followed a different sequence of steps to start fresh).
    \item \textit{Mapping of Privacy Configurations:} To obtain on how many steps are required to disable each app, we mapped the privacy configurations responsible for handling personal data for each app and noted the pathways to each privacy control. We emphasized what privacy control was included in Apple's sources and what was not included. We also present the personal data collected from users' by each app.
\end{enumerate}

\subsection{Results of Study I}
We present a brief summary results of Study I below. Further details are given in the appendices.

%\label{clouddefaultapplications}

During early 2020 we analysed Apple's official documents regarding user's privacy and the data collected to the cloud. We also explored settings on both iPhone (iOS) \textit{version 14.0+} and Macbook (macOS) \textit{version 10.15+}. Choosing these two devices allowed us to find participants who owned at least two different Apple devices due to iPhones' popularity \cite{mobilecosystemsstudy}.
The results of our first study findings are summarised in Tables \ref{table-ppr:privacyconfig} and \ref{tab-ppr:datacollected}. For complete tables please see Appendix \ref{appendixC}. There were several serious issues with information related to privacy and the ways to configure privacy\-related settings.

\begin{figure}[ht]
  \centering
  
  \includegraphics[width=\linewidth]{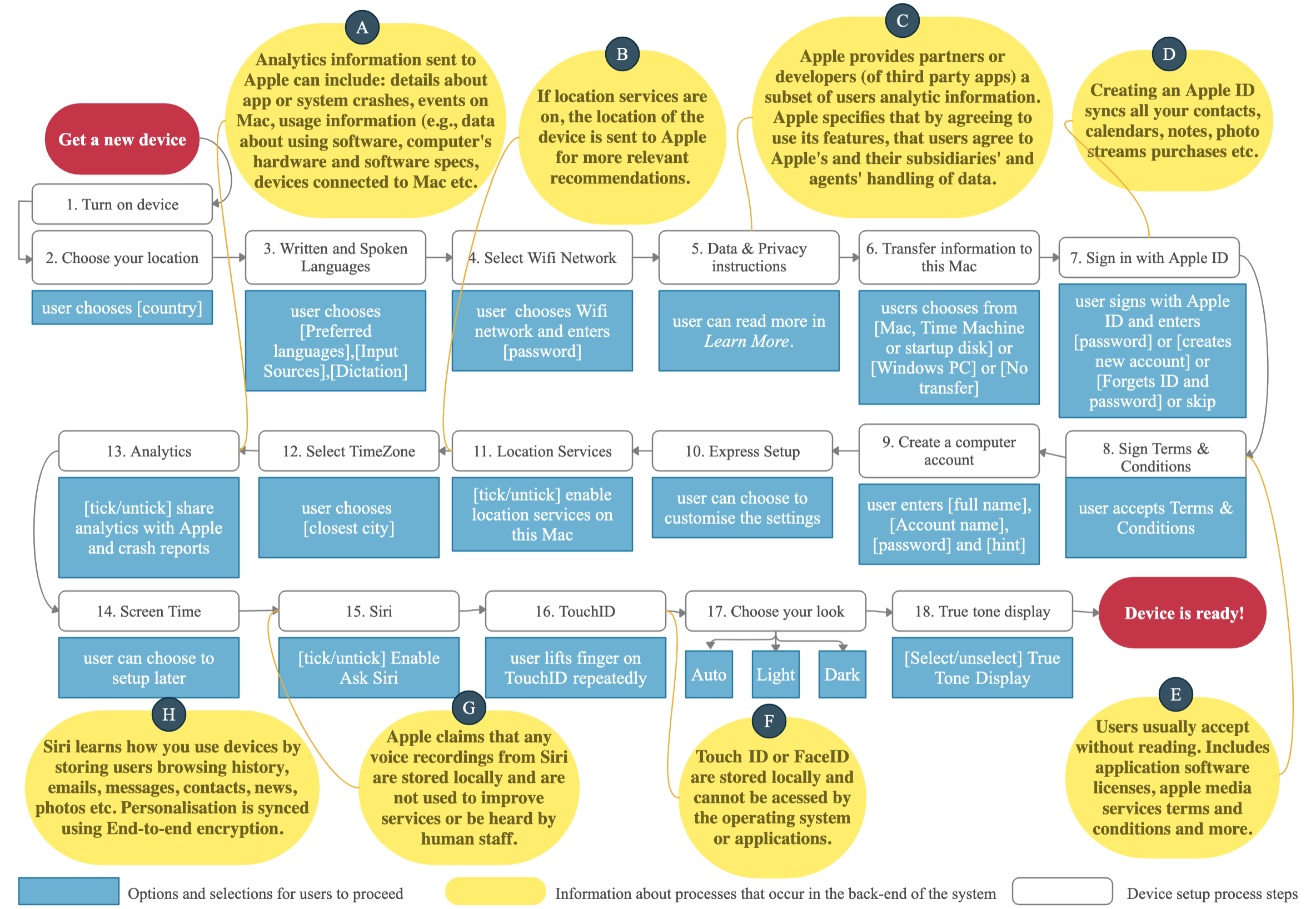}
  \caption{Demonstrates the contrast between the steps users experience and the data handling processes involved at various stages of the device setup process: The user begins the process of setting up their device by purchasing a new device. Steps 1 - 18 explain the steps required for a complete set up of a user's device, for instance a MacBook (macOS 10.15+). Yellow bubbles denoted by letters A - H are captured from Apple's official Privacy Policy statement \cite{Applesprivacypolicy}. Bubbles A - H highlight examples of personal information collection that occurs at various stages of the setup process. In addition to other data handling procedures such as location of the information stored, for instance in F, user's fingerprints are stored locally on the device.}
  \label{fig:back-end-process}
\end{figure}

\textbf{All settings are not documented or are misleading} Apple does not mention all the privacy configurations that are required to be turned off to disable an app from functioning or sharing user's data as shown in Table \ref{tab:privacycontrols1:app} and \ref{tab:privacycontrols2:app} in Appendix \ref{appendixC}. This is also the case when setting up devices for the first time as illustrated in Figure \ref{fig:back-end-process}. In the following, \textit{Siri} as the running example.  

On Apple's official Privacy Policy \cite{Applesprivacypolicy}, the instructions to disable the following privacy configurations of \textit{Siri} can be found: Ask Siri, Siri and Dictation, Siri Personalisation, iCloud Syncing, Transcription. Other \textit{Siri} privacy configurations are not mentioned: Integrated apps, Dictation, Location Services and Request History. 

%This is despite the latter privacy configurations are part of the settings controls of an app responsible for users' data. 

Further, even if an app is supposedly disabled, some settings remain enabled. For example, when first setting up the device, the option to enable \textit{Siri} is disabled. This was done by clicking \textit{Set Up Later in Settings} in the Setup Wizard. Next, when checking the settings of \textit{Siri}, \textit{Siri} was not enabled when first setting up the device; to discover later that \textit{Siri} continues to learn from apps.
    
The information on different configurations is also scattered. For example, disabling \textit{Siri} can be found under \textit{Safari Search} and \textit{Siri and Search} \cite{Applesprivacypolicy}. 
%Moreover, scattered information are observed about disabling certain privacy configurations and its consequences to app functionality on the privacy policy, 

 \textbf{Unknown number of actions required to completely disable data sharing} Shown in Table \ref{tab:privacycontrols1:app} and \ref{tab:privacycontrols2:app} in Appendix \ref{appendixC}, the privacy configurations belonging to an app are described with the respective paths to disable them. Importantly, the privacy configurations marked with an asterisk are the only privacy configurations belonging to an app that are mentioned on the Apple's official documents. No information is found about the remaining privacy configurations.

\textbf{Data handling practices are not disclosed}   Even if users do follow this description and disable the privacy configurations that are provided, users do not receive any confirmation about the previous data they disabled with the app nor are they notified of whether their data is deleted or stored for a longer periods of time. 

%Details of tracing an example case with \textit{Siri} are available in Appendix \ref{AppendixD}. 
Additionally, Apple's documentation is missing information about what personal data actually leaves the device as shown in Table \ref{table-ppr:privacyconfig}. Users' personal data collected by the default apps for reference in Table \ref{tab-ppr:datacollected}.

In summary, we have presented a systematic evaluation of default apps and privacy configurations. We discovered that the features are poorly documented, and have presented summaries in this section and give further details in the appendices. Specifically, we discovered that steps required to disable features of default apps is undocumented and data handling practices are also undisclosed.

 %--------------------------------------------------------------------------
%%====================Study 1 END ============================  

\section{Study II: Qualitative study method}
\label{background}

%================================================================================
%====================== STUDY 2 =================================================
%================================================================================
\label{study2-sec}
To capture users' understanding of privacy configurations of apps within a mobile ecosystem we conducted a qualitative study. %This second study, contributed to exploring users' understanding of privacy configurations and what users' believe they do. As well as their expectation of what occurs when they share their data with the app. We also observed the impacts that setting up these privacy configurations had on users.
\subsection{Recruitment and Screening}

We recruited 15 participants aged 18 or older across Finland. The participants were recruited using the following methods: (1) posts on the university's official LinkedIn page and (2) Facebook paid advertisement for 14 days. %The ad was tailored to reach Apple users who use iPhones and Macbooks who were above 18 years old and who reside in the target country.
We were able to reach a diversity in participants' backgrounds. Following, 67 interested participants completed a survey using Webprobol online survey tool. The screening survey contained questions about participants: owning one or multiple devices, default apps they use, age, occupation, gender and email addresses for contact purposes.

\subsection{Interview Participants}
%%review: why only 15 participants, justify that and include details of participants
We interviewed 5 women and 8 men; 2 participants preferred not to disclose their gender. The majority of our participants were 18 – 29 years old. Participants came from a wide variety of educational and professional backgrounds; Computer Science and IT, Architecture, Business Administration, Art and Design, Industrial Engineering, Economics, Research and Development and unemployed participants. We interviewed participants that have at least used three or more default apps. All participants used Safari, 87\% used Find My and iMessage (No distinction was made between iMessage Cloud service and regular SMS. Both services can be accessed from the same iMessage app.), 80\% used TouchID, Location services, Facetime and the least used apps were \textit{Siri} and Family Sharing as shown in Table \ref{tab:demographics}. Higher usage of default apps is observed in young adults. This also corresponds to recent survey in 2019 \cite{surveyagegroup2019} that estimated the largest consumption of Apple products to exist within young adults between (16 - 24 years old) followed by 25 - 34 years old. A summary of our study user demographics is available in Table \ref{tab:participantsdetails} in Appendix \ref{AppendixE}.

\begin{table}[ht]

  \centering
  \caption{Demographic characteristics of interview participants}
 
    \begin{tabular}{rll}
    \toprule
    \multicolumn{1}{l}{\textbf{Attribute}} & \textbf{Range} & \textbf{Sample Size (N=15)} \\
    \midrule
    \multicolumn{1}{l}{Gender} & Female & 5 (33\%) \\
          & Male  & 8 (53\%) \\
          & Not mentioned & 2 (13\%) \\
    \midrule
    \multicolumn{1}{l}{Age} & 18 - 29 & 8 (53\%) \\
          & 30 - 39 & 5 (33\%) \\
          & 40 - 49 & 1 (7\%) \\
          & 50 - 59 & 1 (7\%) \\
    \midrule
    \multicolumn{1}{l}{Default apps} & Touch ID & 12 (80\%) \\
          & Find My & 12 (80\%) \\
          & \textit{Siri}  & 10 (67\%) \\
          & Safari & 15 (100\%) \\
          & Location Services & 13 (87\%) \\
          & Family Sharing & 4 (27\%) \\
          & iMessage & 13 (87\%) \\
          & Facetime & 11 (73\%) \\
    \bottomrule
    \end{tabular}%
   \label{tab:demographics}%
\end{table}%

\subsection{Ethical Considerations}
Our institution's research ethics committee ruled that this study did not require an ethical approval. In addition, we ensure that we abide by ethical considerations in Computer Science \cite{monlo}: (1) All of our participants were provided information sheets containing information about the study, voluntarily nature of participation and right to discontinue and withdraw participation, data processing and protection details. (2) Before the interview session, we read through the information sheet together with the participant to obtain informed consent.

%% red
\subsection{Limitations}
All studies have some limitations. Due to the qualitative nature of interviews and our constrained (N=15) but diverse sample size, our findings may have limitations to be generalised to and beyond Apple's iOS and macOS. Research suggests that, self-reported answers can provide valuable insights into users' experiences \cite{redmiles:2018:biases}. 
%In our semi-structured interviews, which provide valuable insights when translated with care into real-world contexts . 
Secondly, we focused on this study in two operating systems: iOS and macOS. We recruited participants that owned at least both iOS and macOS devices. This allowed us to reach a wider range of participants. Furthermore, even though our primary focus was to enquire about the use of these two devices, our open-ended questions allowed our participants to express their experiences when using other devices such as iPads, watches and smart TVs. We also observed that participants focused more on the app configurations rather than variations in configurations between devices. A follow up for our work could be a cross-comparison study between different ecosystems.
%% review: explicitly how were the interview questions chosen etc.
\subsection{Interview Sessions}
We conducted semi-structured interviews remotely. The interview sessions took approximately 60 - 90 minutes on average. Participants were compensated with gift cards worth 20 euros after each interview. Interviews were audio-recorded. Teams transcription feature was used to transcribe the audio recordings for further analysis. The transcripts were then proof-read by the first author and checked for any possible syntax issues before analysis.

%% table
% \usepackage{color}
\begin{table*}[ht]
\centering
\caption{Central interview topics, research intentions and sample questions from interview script}
\begin{tabular}{|l|l|l|} 
\hline
\textbf{Topic}     & \textbf{Research Intention}                                                                                          & \textbf{Sample questions from interview script}                                                                                                                              \\ 
\hline
Setup process      & \begin{tabular}[c]{@{}l@{}}To understand how users think \\and process the setup of their devices\end{tabular}       & \begin{tabular}[c]{@{}l@{}}\textit{Did you setup your [device] by yourself? }\\\textit{How was the setup process for you?}\end{tabular}                                      \\ 
\hline
Storage of data    & \begin{tabular}[c]{@{}l@{}}Participants thoughts on where \\their personal~information is stored\end{tabular}        & \begin{tabular}[c]{@{}l@{}}\textit{Where do you think your fingerprint }\\\textit{information is stored?}\end{tabular}                                                       \\ 
\hline
Synchronisation    & \begin{tabular}[c]{@{}l@{}}Participants perception of how their \\personal information is shared\end{tabular}        & \begin{tabular}[c]{@{}l@{}}D\textit{o you share your location on the devices }\\\textit{that have the Family Sharing feature?}\end{tabular}                                  \\ 
\hline
Disabling features & \begin{tabular}[c]{@{}l@{}}To uncover if users are able to stop \\sharing of their personal information\end{tabular} & \begin{tabular}[c]{@{}l@{}}\textit{Do you know how to disable Safari from }\\\textit{sharing your information between your }\\\textit{devices on your iPhone?}\end{tabular}  \\
\hline
\end{tabular}
\label{table:interviewtopics}
\end{table*}

\subsection{Qualitative Data Analysis}
Topics for the interviews were formulated after reviewing previous work on the topic of privacy configurations in apps \cite{Ramokapane2019,smullen:2020}. The interview script was then designed to include questions inspired from the previous work and questions suitable to the scope of this work. Central topics discussed (also see Table \ref{table:interviewtopics}): (a) setup process of devices, (b) storage of personal information, (c) synchronisation of personal information between devices and (d) the ability to disable information sharing between devices through apps. The interviews structure was divided into two parts: general questions and tasks. For the first part, participants were not informed on how to disable features or how they actually work as we aimed to gather their own understanding of the functionalities of default apps. We probed the participants about setting up their devices and perceptions of the process. Following in the second part (tasks), we asked the participants to try to disable certain features on MacBook and iPhone. We have also asked participants about their perceptions of what happened after they disabled certain features. After interviewing 10 participants, we arrived at a point of theoretical saturation. We resumed with 5 more participants to validate the saturation stage. The interview phase ended with 15 participants. The interviews resulted in 925 minutes of audio and 709 pages of transcripts.

To analyze and code the interview data, we used a hybrid approach: initial a-priori coding followed by inductive coding. Our coding process is one of many approaches common in HCI for reliability of data analysis \cite{howmanyinterviewsareenough:2006}. The first author developed a codebook based on high-level categories, which included user experiences when setting up their devices, their knowledge of different app functionalities and disabling features, and what they know about the handling processes of their personal data by the default apps. Next, the first author coded two interviews under the high-level categories adding inductive codes. The first and second author then discussed the codebook to resolve any disagreements. The following step was to use the updated codebook to code two more different interviews. The two authors met again for a discussion to agree on a common codebook. Once that was achieved, the coding of the remaining interviews was completed by the first author. We identified and organized categories as well as finding relationships in our data. The results of our data analysis are found in the next sections.

\section{Results of Study II}

%================================================================================
%====================== RESULTS - STUDY 2 =================================================
%================================================================================

In this section, we present the results of our semi-structured interviews. We grouped our findings into three main themes that explain users understandings of the privacy configurations and the impacts of making related decisions. The three themes are: (1) Understandings of Configurations (2) Structure of Privacy Configurations and (3) Impacts of Setting Privacy Configurations. In addition to these three themes, the section \textit{Verification: Challenges and Misunderstandings} shows the results from tasks asked from users to perform. The tasks contributed to confirming users challenges and misunderstandings about the mobile ecosystem.

Our results show that our participants were surprised that some of their information was shared by default. Participants were not able to disable default features. We also present unexpected results: participants were also aware that they were being tracked even though were surprised about the extent of it; participants were also aware that stopping data sharing does not guarantee that it is not shared anymore; tracking introduced trust issues in family relationships.

Our findings further imply that several factors were found to contribute to privacy decisions of apps. These factors include being aware that default apps were tracking users as well as their awareness of data retention. The structure of privacy configurations also contributed to the making of privacy decisions. This includes the level of clarity of privacy configurations, transparency and levels of organisation. Overall, setting privacy configurations had several impacts on users such as causing a misunderstanding of functionalities, inability to disable them and finally contribute negatively to relationships. 

%We present the results in two sections below. First, we present the results of the semi-structured interviews. Then, we present results of specific tasks that we asked the participants to conduct.

\subsection{Theme 1: Understandings of Configurations}
Our analysis revealed perceptions that users have of privacy configurations regarding different default apps. Several participants believed they were being tracked through some default applications. Others were aware that stopping data sharing does not guarantee that their data is not shared anymore. Here data sharing can imply sharing with the service provider (Apple) or within the devices in an ecosystem.

%=====================================
\subsubsection{Awareness of Tracking} 
%=====================================
11 out of 15 participants (P01 – P04, P06, P08 – P13) believed they were being tracked when using Location services, Siri and Safari. We have collected these answers from participants prior to sharing information related to default apps. Participants expressed that they knew that they were being tracked while using the mobile ecosystem. Participants elaborated that tracking is useful sometimes but can also be invading. They shared their experiences with default apps tracking them and how that impacted the situation; Participants thought that at certain instances, the tracking was useful to locate lost devices. Other times, it was rather invading. Participants expressed their awareness that Location services, Siri and Safari were tracking them.

%locations services

Participants explained that they are aware that different apps using location services are tracking them. P01 explained that when using Maps on airplane mode that she cannot be tracked anymore, then realized that in fact her location was still traceable. She explained: \textit{``I thought previously that when I turned airplane mode on, they could not track me anymore. But now like I was walking around a [city] the other day with airplane mode and like it could still track me down every street.''} 

P01, P04 and P09 thought tracking on Find My can be useful to find missing persons. P04 shared instances where Find My could be helpful to track people: \textit{``You could use it to find missing persons as well or some kind of emergency like that.''}; P09 shared a story that recently happened to him when hitchhiking in a city and losing his phone as well his family members and being able to track them: \textit{``they had stopped to a MacDonald’s so they didn't hear with their thing ringing .. I got my phone back but yeah it was great because I didn't have any of their contact information otherwise.''}

P02 had a different take on Apple overriding location permissions to assist users in tracking lost devices in cases of emergency describing it as a "two edge sword": \textit{``It shouldn’t be on at all times because there are also situations when someone does not want to be found. It is a little bit two edge sword.''}

P10 had several concerns about tracking related to Siri tracking her searches: \textit{``It hasn't only been the voice, but it also does with the search. Like when you start to search for something .. the app suggestions are mostly quite accurate, but the other things are not like the send an email to or contact this person or your album or whatever. I don't think those are accurate and I don't like it.”}; P01, P10 and P11 also expressed their concern about Siri tracking users; P10 shared concerns related to personalizing ads for users which has caused the content to be very tailored and restricted \textit{“I feel like I get targeted .. I don't want to be in a bubble where I see only things meant for a 23 year old young woman. I want to see more than that.”}; P11 shared when asked if they thought Siri collected more that it needed to track users: \textit{“Every company knows more than my mother these days.”}; P03 was concerned if turning off sharing in Safari guarantees that data is not shared anymore between their devices therefore enabling tracking: \textit{“In the end every device can be tracked independently.”}

%=====================================
\subsubsection{Awareness of Data Retention}
%=====================================
9 out of 15 participants (P01 – P04, P06, P10, P11, P14, P15) were aware stopping data sharing does not guarantee it is not shared anymore. We asked the participants about their thoughts about data retention prior to telling them information about it. Participants were aware that data sharing was not completely disabled and data was not completely deleted from Safari, Siri and Location Services. 

%safari

P01 expressed that it cannot be guaranteed that data collection is completely disabled: \textit{“I can't knock on Apple store to be like hey, did you actually do this but are better to have the possibility of it being disabled then? Knowing 100\% that's not disabled?”}. On the other hand, P03 was not very worried about not being able to completely delete information: \textit{“.. but since we anyways have our accounts linked to this Apple ID's .. I don't really know whether that information can be stored somewhere else.”}

%siri

P02 explained what he thought happened to his data in the long run: \textit{“they probably have all my data stored for more than two years.”}; P04 when attempting to disable Siri from sharing information: \textit{“It shouldn't be listening what you do. And then it's not shown on the menu bar and and it won't react to your hey Siri or whatever command you have on for accessing Siri .. at the same time, we don't know whether it's actually turned off, or whether Siri is listening..”}

%location services

P02 compared disabling Safari from sharing his personal information: \textit{“But what bothers me is just like in an iPhone you can turn off your location sharing with the family with a click. While it's a lot harder on the MacBook.”} Then he continues to share his thoughts about completely disabling data sharing: \textit{“.. I think it just stops sharing with the persons that I stop sharing with but it doesn't stop sharing with Apple.”}

%imessage and facetime

P02 talked about the inability to delete messages in iMessage from where they are really stored: \textit{“The impression that I can delete it from my device and not from any kind of storage where they stored it..”}; P06 similarly agreed: “Like it would mean that they could read it straight from my phone, but not in between.”

\subsection{Theme 2: Structure of Privacy Configurations}
%=====================================
\subsubsection{Transparency as a requirement}
%=====================================
Participants want to know what happens to their information. 11 out of 15 participants (P01, P02, P04 – P06, P11, P12, P15) want more transparency from the ecosystem. They want to know what privacy configurations are enabled by default and the data being stored. Participants thought that more transparency in instructions can be useful. They also explained that privacy configurations can be too scattered.

P01 explained that there should be a balance in providing information related to configuring privacy settings: \textit{“You want that balance, right? because you'll have people who are very ignorant and they don't think about this .. that information should be given to them.”}

P02 shared on needing transparency with regards to data handling: \textit{“Someone else is selling my data without my knowing or with my knowing .. people want to know what is happening to the data.”}; P02 also described privacy configurations to be “cloudy” and scattered: \textit{“Because I do think that iCloud settings are a little bit kind of cloudy about what they actually do.”} Further elaborating on Siri, \textit{“I think it's quite fine with them collecting my personal data to improve experience. But I do think that it should be a little bit more transparent about what kind of data they are collecting to improve my experience.”} 

P02, P04 and P12 agreed that users need more information about data handling practices from mobile ecosystem service providers. P04 was asked if Apple were transparent with their users: \textit{“Every instance can be more transparent with the stuff nowadays. Especially the ones [meaning leading mobile companies] aiming to have some financial advantage over others to make money.”} He also added: \textit{“It’s a good question whether a company needs to store everything for long periods of time. What they actually do with all the data for transparency’s sake. Maybe also give some kind of timeline for the information storage.”}.

%=====================================
\subsubsection{Organization of Privacy Settings}
%=====================================
4 out of 15 participants (P01 – P03 and P11) thought that privacy configurations of default apps are not organized in a clear way. In fact, it was difficult to find certain settings at times. This had direct effects on being able to configure these privacy settings or re-configure them later. For example, P01 attempted to change the owner of the second-hand iPhone she received. She explained that she thought it was an intuitive process but it turned to be otherwise: \textit{“.. I kept trying to change the owner of the phone. So like I think it's the associated Apple ID. It still kept defaulting to my sister's and my sister was getting all of the account notifications on things that were supposed to come to me. I thought it was intuitive, but it clearly didn’t work.”}

P03 explained that the Terms and Conditions were lengthy: \textit{“I don’t really know. Sometimes you simply don’t have time to go through all the policies and stuff. I don’t think it is clear enough.”}

P02 thought that it was confusing that System Preferences contains security settings as well as iCloud on MacBook: \textit{“In system preference on the MacBook, we've got Apple ID and then we've got iCloud, but then we also got Security and privacy as one option. It's very scattered.”}

%=====================================
\subsection{Theme 3: Impacts of Setting Privacy Configurations}
%=====================================
\subsubsection{Surprised Users}
%=====================================
We asked the participants about what surprised them when using the mobile ecosystem. These questions were asked after we told the participants information about default apps. Participants were surprised that Siri, Safari and Location Services enabled certain data sharing by default. 9 out of 15 participants (P02, P06 - P09, P11, P14, P15) expressed that they were surprised that these privacy configurations did not match what they really wanted. Participants shared their experiences when navigating the privacy settings of Siri, Safari and Location Services.

%\textbf{Siri}
Participants were surprised that Apple learns from apps. We asked the participants who used Siri if they were aware that she learns from apps to make suggestions across the app. Participants were surprised that this was enabled by default.

P11 accessed the apps that were reporting app data back to Siri: \textit{“I am a little surprised to see that it's collecting data from all apps in iPhone.” } 

%P09 shared the same reaction when seeing the list of apps that report back to Siri: \textit{“.. Why is it that Siri gets all that data and not just straight up Apple?”}

Participants were surprised to learn that small subsets of requests that have been reviewed from Siri may be kept for more than two years. P08 shared about Apple reviewing Siri recordings without a random identifier: \textit{“That is interesting, very nefarious actually .. Super shady.”}

%\textbf{Safari}
Participants were surprised to know Safari shares data to Apple. We asked the participants who used Safari if they were surprised to learn that Safari may send browsing information such as location, topics of interest, search queries and suggestions to Apple. Participants were surprised that Safari was reporting back to Apple by default. P06 on Safari reporting data back to Apple: \textit{“Uh I wasn’t aware of that .. if I would search for anything sensitive, then I would switch browser.”}

%\textbf{Location Services}
Participants were surprised to see the list of apps that had access to their location. We asked the participants who used Location Services if they were surprised to see the list of apps that had access to their location. Participants were surprised to see some apps that were using the location information by default.

P14 was shocked to see the list of apps using his location especially photos app: \textit{“I’m shocked. I have never with photos [referring to never providing permissions for photos to access geolocation]. That should be while using the app.” }
%=====================================
\subsubsection{Tensions in relationships}
%=====================================
Family sharing app was considered useful and invading. We asked the participants that have used Family Sharing about their experiences. 7 out of 15 participants (P01, P02, P03, P06, P07, P14, P15) thought Family Sharing as a default app can be useful. Family Sharing enables members to share their location, purchases and other useful things. Participants believed this can introduce distrust to families and relationships because of the monitoring. Participants found family sharing to be useful, and in other instances rather invading.

%\textbf{Family Sharing Useful}\\
P02, P07 and P013 found Family Sharing to be useful for multiple purposes; for example, safety or saving money. P02 was able to save money with Family Sharing: \textit{“I think it's just kind of saved money in some way just by sharing, because you can share some purchases and stuff like that.”}

Other participants were concerned about safety, P07 has a child. She would like her to know exactly where she is: \textit{“I want her [child] to be able to know where I am. I share my location with her.”} 
%P13 who also has children explained: \textit{“I have two teenagers .. so I always know at least the last known location of them.”}

%\textbf{Family Sharing Invading}\\
P01, P13 and P15 found Family Sharing to introduce issues to family relationships. They shared their experiences on how they thought monitoring other family members can cause trust issues. Participants thought that at times, Family Sharing can be invading to family relationships. P01 about Family Sharing feature and how it can be invading of teens privacy: \textit{“Maybe too much like parents going too far to track their child when they are about 17 or 18 years old.”}; P01 also shared her thoughts on tracking minors using Family Sharing: \textit{“Point is that I know it has caused a lot of tensions in new avenues of how families and friendship tensions can develop and manifest.”}

P13 explained accidental sharing that could occur when using Family Sharing leading to tensions: \textit{“Sharing a photo that they would not want to share with each other, that goes to a wrong sort of similar to thing with messages that every now and then everyone sends a message to wrong group which may cause difficulties.”}

\subsection{Verification: Challenges and Misunderstandings}
In the second part of the interviews, we have verified users knowledge of privacy configurations of different default apps. In this section, we include the challenges and misunderstanding that users have about the eight default apps: TouchID, Family Sharing, Siri, Safari, Find My, iMessage and Facetime and Location Services. %An example is show in Figure \ref{fig:cross-comparison-siri} of a participant attempting to disable Siri.
%=====================================
\subsubsection{Understanding of Functionality}
Participants were confused about what happened to their information. We asked the participants about their understanding of the functionalities of the default apps. We collected their answers during the tasks. All participants were unsure what privacy configurations did and how the ecosystem works in Siri, Location Services, iMessage and Facetime and Safari. 

%\textbf{Siri} \\
Participants P04 and P07 were not sure where Siri stores the voice recordings. P04 was unsure if enabling Siri meant that Siri was storing his voice recordings when making queries: \textit{“I would guess that all the voice being recorded is actually being recorded and stored somewhere for development reasons and making the AI better as well as the voice recognition better with different languages.”}; P04 and P07 shared concerns about disabling Siri: \textit{“And then it's not shown on the menu bar and it won't react to your Hey Siri or whatever command you have on for accessing Siri, we don't know whether it's actually turned off, or whether Siri is listening.”}; P07 was sure that Siri was listening all the time: \textit{“I believe all devices that are allowed to listen are collecting more than when you use it.”}

%%\textbf{Location Services}
%P04 and P10
 Participants were unsure if disabling location services meant that their location cannot be shared anymore with the ecosystem or otherwise. We asked the participants about what they thought about their location data and what happens if they disable it. P04 shared his experience disabling apps from accessing location services that they cannot find his location anymore: \textit{“The phone itself has GPS operations going on anyway. Even when I have the data center off. Of course you could find a location for the iPhone device.”};

%\textbf{iMessage and Facetime}

Participants were unsure what goes on in iMessage and Facetime. P01, P10 and P11 expressed that they were confused about the different menu on iPhone and MacBook which was confusing to configure. P10 about iMessage and Facetime using End-to-End encryption: \textit{“If that's true, then why do criminals and people like that still use like the the dark web and Telegram, and those kind of things if it's true?”}

P11 was trying to disable iMessage and Facetime on his iPhone. When trying on his Macbook: \textit{“I was just going to say that with Apple it's nice work, same way in different places, but not at this time.”} He tries to access it from the iCloud then explains in confusion: \textit{“The not in iCloud .. it has its own settings. Yeah, that's funny, right? Because it's .. the messages are in iCloud, but that's .. little bit illogical.”}

P01 was confused to learn that iMessage in her iPhone was synced to the one in her MacBook even though she has never used it before on her MacBook: \textit{“It's definitely synced up with my phone. That's not good.”}

%%\textbf{Safari}

Participants P02, P06, P05, P09, P11, P13 were unsure what happens when disabling Safari. We asked the participants what they thought happened with Safari and where their data was going. Participants expressed that when they delete information from one device that it should remove from all devices. Participants were also confused about if deleting History, Bookmarks and Cache were enough to disable sharing.

P13 was confused about when happens to his data when he tried to disable Safari from Sharing information to other devices: \textit{“so say that it's lost and I want to this to be erased. I believe that it works, so only that it removes the device and .. Uh, iCloud, but not the individual devices .. I'm actually not sure how it works.”}; P09 tried to disable Safari from sharing information: \textit{“If I didn't have Google, it would be impossible to find that.”}; P05 similarly agreed: \textit{“.. I don't think it would completely erase it like I couldn't probably get to it .. but I think there would be like some traces on it from it.”}; P02 was unsure about the same task: \textit{“Uh, no I’m not sure about it.”}; P11 attempted to disable Safari bookmarks: \textit{“I'm not sure that if I disable the safari bookmarks, does it stop sharing everything?”}; P06 attempted to disable Safari from sharing information between devices on MacBook: \textit{“Uh, and I'm not sure what would happen for the data that is already synchronized. I think it just stays there, 'cause I guess there is like a common database ..”}

%=====================================
\subsubsection{Disabling data sharing}
%=====================================
All 15 Participants did not know how to disable data sharing on Safari, Siri, Location Services, iMessage, Facetime, and Family Sharing. Participants struggled with disabling data sharing in default apps. Participants shared their experiences while attempting to disable the previous default apps from sharing their information. 

%\textbf{Safari}
    
Participants did not know how to disable Safari completely from sharing personal information on their devices. We asked the participants to disable data sharing in Safari. Participants were unable to completely stop data sharing in Safari.

P01 tried to disable Safari from sharing her data on her iPhone. She first went to Settings > Privacy and Security > Advanced. She shares: \textit{“I'm trying to find if there's like a specific like security option on the iPhone.”} She goes back to Settings and navigates: \textit{“so I'm back in the settings page and I'm trying to find .. So earlier I was under Safari app. Now I'm trying to find like the general iPhone .. Security is what I'm intuitively trying to do..”} She finally shares her final assumption: \textit{“I have no idea. This is where I would know I don't know.”} P15 tried to disable Safari from sharing her data on her iPhone. She first went to Settings > Safari > Clear History and Website data. She explained that she was not sure that this disabled. She then tries on her MacBook and expressed her struggle to disable bookmarks: \textit{“For bookmarks I’m having more trouble. I'm just thinking, how complicated can it be to delete the bookmark even I'm even trying to.”}

%\textbf{Siri}

Participants were not able to disable Siri completely from collecting information from apps on their devices. We asked the participants to disable data sharing in Siri. Participants were unable to completely stop data sharing in Siri. P05 tried to disable Siri. She tries to access on her iPhone, General > Siri \& Search > Suggestions on Lock Screen. Then she navigates the apps under Siri sub-menu and shares: \textit{“Then there's press home for Siri and allow Siri. When locked, there are these trade toggles. I think that if I were to toggle them off, Siri would not show up, I think.”} P07 when trying to disable Siri from the MacBook. She shares: \textit{“Alright, just couldn’t find it.”}

%P11 when trying to disable Siri from his iPhone. He goes to Settings > Siri \& Search. He explains: \textit{“Yeah, so I. I think this is the list of all apps here in under the Siri and search and you can.”}

%\textbf{Location Services}

Participants were not able to disable their location data from being shared completely. P08 tries to disable the apps that are collecting his location data on his iPhone. He shares: \textit{“I'm surprised that I can't find it. I would have thought there would be a location services tab here and I'm going to General. Now I'm going back to settings. Let’s open iCloud. No It’s not there either. How do I turn off location services? I think I would have to Google this on my phone.”}

%\textbf{iMessage and Facetime}\\
Participants were not able to disable iMessage and Facetime features completely. Participants were unaware how to disable iMessage and Facetime or modify their privacy configurations. P05 tried to disable iMessage and Facetime from her MacBook. She shared: \textit{“Well, this is made a little trickier here on the MacBook.” She heads to System Preferences: “quite more labor intensive to at least find it.”} P08 tries to disable iMessage from his MacBook. He shares: \textit{“So I can sign out of iCloud and then it gets rid of that. But there is no way of like disabling it because that's the only way of sending messages as I understand.”}.

%\textbf{Family Sharing}\\
Participants were not able to disable family sharing features completely from sharing their personal data. P15 shares when trying to disable Family Sharing from public sharing: \textit{“I’m going to Apple ID doing family sharing. It seems like my location is not shared. It's asking me to share it so it must not be shared. I couldn’t find anything.”}; P07 tries to disable Family Sharing from her MacBook. She shares: \textit{“I have no idea where to find the iCloud.”} Then she accesses System Preferences > Apple ID > Family Sharing.

%-------------------------------------------------------------------------------
\section{Discussion}
\label{discussion}
%-------------------------------------------------------------------------------
%%%%
%================================================================================
%====================== STUDY 2 - DISCUSSION=============================================
%================================================================================

\label{discussion-sec}

We have presented two studies that show the challenges on configuring desired privacy settings in the Apple iCloud mobile ecosystem.
Our first study's findings show that the instructions to configure the privacy of default apps are not straightforward. Moreover, the required steps to disable privacy configurations are unclear. Upon attempting to disable certain privacy configurations, limited confirmation messages are prompted for users. We were interested whether users' perceptions align with what we have observed in Study 1. Our results of Study 2 describe that users do not understand data sharing between default apps. Users were confused whether disabling a feature meant that their personal information was not shared anymore. Our tasks verified that users were in unable to disable any of the default apps. Furthermore, after users later learned how their information was handled they were surprised. Users denoted the challenges they faced locating and disabling default apps to the unclear organization of privacy settings of default apps. Finally, users talked about possible tensions arising in family relationships due to information sharing through default apps.

Next, we will discuss our findings for the specific research questions we set to answer.

\paragraph{RQ1: What privacy configurations are available to control default apps? }

%% clear and short sentences. the privacy policies or instructions are ambiguious. not informative. (check privayc policy work). not clear why it is the second paragraph.
Data handling in a mobile ecosystem is ambiguous to users. Through our system evaluation in the first study, we were able to deduce a level of ambiguity associated with handling personal information in a mobile ecosystem. For instance, many instructions related to data handling of default apps do not contain enough description. For instance using phrases like ``a small subset of requests may be kept to beyond two years'' and ``Siri data'' \cite{siriprivacypolicy} does not include proper explanation of what this data is or which subset of data is specified. Users have also expressed eavesdropping concerns specifically towards the voice assistant \textit{Siri}. A study performed on smart speakers supports our results that participants were not aware that their personal data was stored for long periods of time \cite{Malkin:2019:smartspeakers}.

% the amount of steps required ranges from N steps. put things to context. what does this mean to the users. 
Privacy configurations of default apps require multiple steps to disable certain features. The amount of steps required to completely disable privacy features depend on the app (see Appendix \ref{appendixC}). Having multiple steps required to disable app features is confusing and time-consuming for users. This is especially a problem when there are no confirmation messages that certain features are completely disabled. %Or that data has been completely removed.

%% Review: answering the research question 1
\paragraph{RQ2: How can users control default apps?}
%``something quoted here.''
% what was the point of study 1. 
The information provided users on how to configure the privacy of their personal information is ambiguous. The ambiguity can be denoted for vague descriptions of controlling privacy configurations for example, missing steps to configure default apps (see Tables \ref{tab:privacycontrols1:app} and \ref{tab:privacycontrols2:app} in Appendix \ref{appendixC}).

During initial device setup, users are prompted with options to disable or enable default apps. Descriptions about the data handling practices of default apps are not easily visible next to the check boxes to enable these apps. Descriptions such as how data is to be handled, what type of data is collected, how to disable or enable privacy configurations, are not found when users' setup their devices. Descriptions of data handling practices are typically present in Privacy Policies of different default apps (available to users on the web, often via embedded links in the app settings) \cite{siriprivacypolicy,locationservicesppolicy,iMessage:Wiki}.

We suggest that not presenting descriptions of privacy controls during setup leads to many issues. Users are not able to make informed privacy choices if the information related to their privacy is not easily visible to them.
We observed that users were not able to easily locate the privacy configurations after device setup. Presenting adequate descriptions of privacy controls is needed. Past work has shown that users do not change settings they initially setup, under the illusion that they are recommendations from the platform \cite{Acquisti2015PrivacyAH, Ramokapane2019}.

During the study, we observed that it was difficult for participants to find privacy configurations for default apps on their devices. Participants were not able to disable features of default apps as part of the practical tasks. Participants would perform one or two steps in an attempt to disable a certain feature or app altogether, only to express confusion about what actually happened. When navigating the privacy controls of default apps, participants that they were surprised certain privacy configurations were enabled. Further expressing that they did not recall enabling a certain configuration when setting up the device.

%-------------------------------------------------------------------------------
\paragraph{RQ3: How do users understand privacy configurations and their privacy and security implications?}

Our results show that that participants did not understand how Apple's mobile cloud ecosystem works. Participants were not able to distinguish between what information is shared within the mobile ecosystem (device to device) and between the ecosystem and the service provider (Apple). This confusion led some users to use their own techniques to protect their information, for example, switching browsers (from \textit{Safari} to another browser). Prior work suggests that users tend to leave default feature configurations unchanged if they are confused about what the configurations do \cite{conti:2010}. 

Users did not understand what data is being collected and shared. Additionally, participants did not know what data is stored on each device. These findings align with previous work that suggests that user's do not understand data sharing processes \cite{kang2015dataeverywhere,Ramokapane2019}.

Our findings revealed that most of the participants assumed they were being tracked when using different apps within the mobile ecosystem. Our findings support previous work that suggested that users are aware of Location Services being turned on due to the requests they receive from apps \cite{Consolvo2005locationdisclosure,Ramokapane2019}.

Many participants (9/15) were aware that disabling data sharing does not guarantee that it is not shared anymore to the cloud. Some participants thought that this is likely due to their accounts being linked to the same cloud account.

\label{impactsofprivacyconfigu}
%\subsection{Impacts of privacy configurations}
%-------------------------------------------------------------------------------tensions and susprised users
\paragraph{RQ4: How does setting up default features impact privacy of users?}

Our results showed that privacy configurations of apps such as \textit{Family Sharing} can introduce tensions in family relationships. The distrust and tension may also be attributed to various social contexts and norms that technologies operate in. For example, in some countries cultural beliefs and practices dictate that women are expected to share their devices with other members of the family \cite{Sambasivan:2018:richwomen}.

Despite the potential negative aspects of \textit{Family Sharing}, many participants thought \textit{Family Sharing} app was helpful in  maintaining safety for children through location tracking. With regards to information collected from users by \textit{Family Sharing}, participants felt confused. They were confused as to what why specific personal information were shared between family members who use \textit{Family Sharing}, for instance, sharing purchasing information.

During the interviews, users were surprised to learn about data handling practices of default apps. Participants were surprised that the apps learn from other apps in order to make personalized suggestions. In other occasions, participants were surprised to learn that their personal data may be kept for even two years \cite{siriprivacypolicy}. Our findings align with previous literature suggesting that users misconceptions related to data sharing can cause users to feel surprised \cite{Balbeko2013LittleBrothersWatchingYou} and in other contexts confused \cite{joshuaexplanationforpermission2014}.

\paragraph{Recommendations}

To address the privacy issues with default apps, we recommend the following.

Privacy configurations should be accessible from one location in the platforms. In contrast, currently on macOS, \textit{Safari} settings are accessed from at least three different locations: i) the app itself, ii) iCloud settings, and iii) Security and Privacy. We suggest that the current structure of privacy and security settings of default apps is ineffective and confuses users. %To further improve usability, we recommend improved settings layout that eases the search process of different configurations. 

The platforms should increase transparency regarding information collected and how it is handled. Our participants expressed that they were confused about how their data handled. Previous work has reported users find Privacy Policies challenging in many ways, which impacts their knowledge of how data is handled \cite{Alohaly2016quantificationprivacypolicies,Coen2016paradox,Gamba:2020:preinstalledapps,Kelley2012conundrumofpermissions}. In our study, participants were unable to clearly answer questions related to what happens to their data when they alter a specific configuration. For instance, when participants attempted to disable default apps, participants were not sure what they thought happened to their data. Participants expressed that they did not remember allowing a default app to do what was enabled. Prior work has also suggested that displaying controls early on during the lifetime of a device might not be transparent enough for users \cite{Ramokapane2019}. We encourage vendors to provide illustrations or simple graphics to simplify how users' data is shared within the ecosystem before agreeing to use a default app.

As an example of increasing transparency, vendors must clarify what personal information is required to go through the cloud for a service to work. We observed a level of ambiguity related to users’ knowledge about configuring their default apps. We argue that it is necessary that users know more about their privacy when using default apps. This is important because generally users are not able to uninstall default apps (uninstalling requires root privileges \cite{Gamba:2020:preinstalledapps}).

%As default apps can be described to be sticky – often requiring root privileges to be uninstalled and are not available on the App or Play store for monitoring for their data collection practices like other apps \cite{Gamba:2020:preinstalledapps}.

To further improve users' awareness of privacy controls, we suggest periodic prompts for privacy controls. These prompts should be aimed to assist users in understanding different ways that they control their privacy in the platform. Prompts could also explain how data is handled by the ecosystem and how users can control different aspects of their data sharing process. %Furthermore,  educating users on privacy and security mechanisms available to maintain their privacy online \cite{smartphoneprivacy:2022:Frik}. }

\section{Conclusions}

We found that the seamless integration of smart devices with the cloud comes at the expense of users' privacy.
Our work shows that users may disable default apps, only to discover later that the settings do not match their initial preference. In this paper, we presented two studies. First, we evaluated the privacy configurations of default apps. Second, we conducted interviews to understand users’ perceptions of these privacy configurations. Our results clearly demonstrate users are not correctly able to configure desired privacy settings of default apps. We discovered that some default app configurations can even have negative impact on trust in family relationships.

\newpage


\newpage
\bibliographystyle{ACM-Reference-Format}
\bibliography{citations}


\begin{thebibliography}{47}

%%% the Web address.

\ifx \showCODEN    \undefined \def \showCODEN     #1{\unskip}     \fi
\ifx \showDOI      \undefined \def \showDOI       #1{#1}\fi
\ifx \showISBNx    \undefined \def \showISBNx     #1{\unskip}     \fi
\ifx \showISBNxiii \undefined \def \showISBNxiii  #1{\unskip}     \fi
\ifx \showISSN     \undefined \def \showISSN      #1{\unskip}     \fi
\ifx \showLCCN     \undefined \def \showLCCN      #1{\unskip}     \fi
\ifx \shownote     \undefined \def \shownote      #1{#1}          \fi
\ifx \showarticletitle \undefined \def \showarticletitle #1{#1}   \fi
\ifx \showURL      \undefined \def \showURL       {\relax}        \fi
% invisible to TeX
\providecommand\bibfield[2]{#2}
\providecommand\bibinfo[2]{#2}
\providecommand\natexlab[1]{#1}
\providecommand\showeprint[2][]{arXiv:#2}

\bibitem[Acquisti et~al\mbox{.}(2015)]%
        {Acquisti2015PrivacyAH}
\bibfield{author}{\bibinfo{person}{Alessandro Acquisti}, \bibinfo{person}{Laura
  Brandimarte}, {and} \bibinfo{person}{George Loewenstein}.}
  \bibinfo{year}{2015}\natexlab{}.
\newblock \showarticletitle{Privacy and human behavior in the age of
  information}.
\newblock \bibinfo{journal}{\emph{Science}}  \bibinfo{volume}{347}
  (\bibinfo{year}{2015}), \bibinfo{pages}{509 -- 514}.
\newblock


\bibitem[Alohaly and Takabi(2016)]%
        {Alohaly2016quantificationprivacypolicies}
\bibfield{author}{\bibinfo{person}{Manar Alohaly} {and} \bibinfo{person}{Hassan
  Takabi}.} \bibinfo{year}{2016}\natexlab{}.
\newblock \showarticletitle{Better Privacy Indicators: A New Approach to
  Quantification of Privacy Policies}. In \bibinfo{booktitle}{\emph{Twelfth
  Symposium on Usable Privacy and Security (SOUPS 2016)}}.
  \bibinfo{publisher}{USENIX Association}, \bibinfo{address}{Denver, CO},
  \bibinfo{numpages}{7}~pages.
\newblock


\bibitem[Apple(2021)]%
        {Applesprivacypolicy}
\bibfield{author}{\bibinfo{person}{Apple}.} \bibinfo{year}{2021}\natexlab{}.
\newblock \bibinfo{booktitle}{\emph{Apple Privacy Policy}}.
\newblock
\urldef\tempurl%
\url{https://www.apple.com/legal/privacy/en-ww/}
\showURL{%
Retrieved September 9, 2021 from \tempurl}
\newblock
\shownote{Web link}.


\bibitem[Apple(2022a)]%
        {applecosystem}
\bibfield{author}{\bibinfo{person}{Apple}.} \bibinfo{year}{2022}\natexlab{a}.
\newblock \bibinfo{booktitle}{\emph{Apple Ecosystem}}.
\newblock
\urldef\tempurl%
\url{https://www.apple.com/privacy/docs/Building\_
  a\_Trusted\_Ecosystem\_for\_Millions\_of\_Apps.pdf}
\showURL{%
Retrieved February 18, 2022 from \tempurl}
\newblock
\shownote{Web link}.


\bibitem[Apple(2022b)]%
        {siriprivacypolicy}
\bibfield{author}{\bibinfo{person}{Apple}.} \bibinfo{year}{2022}\natexlab{b}.
\newblock \bibinfo{booktitle}{\emph{Ask Siri, Dictation \& Privacy}}.
\newblock
\urldef\tempurl%
\url{https://www.apple.com/legal/privacy/data/en/ask-siri-dictation/}
\showURL{%
Retrieved January 13, 2022 from \tempurl}
\newblock
\shownote{Web link}.


\bibitem[Apple(2022c)]%
        {locationservicesppolicy}
\bibfield{author}{\bibinfo{person}{Apple}.} \bibinfo{year}{2022}\natexlab{c}.
\newblock \bibinfo{booktitle}{\emph{Location Services \& Privacy}}.
\newblock
\urldef\tempurl%
\url{https://www.apple.com/legal/privacy/data/en/location-services/}
\showURL{%
Retrieved February 16, 2022 from \tempurl}
\newblock
\shownote{Web link}.


\bibitem[Bailey et~al\mbox{.}(2012)]%
        {monlo}
\bibfield{author}{\bibinfo{person}{Michael Bailey}, \bibinfo{person}{David
  Dittrich}, \bibinfo{person}{Erin Kenneally}, {and} \bibinfo{person}{Doug
  Maughan}.} \bibinfo{year}{2012}\natexlab{}.
\newblock \showarticletitle{The Menlo Report}.
\newblock \bibinfo{journal}{\emph{IEEE Security \& Privacy}}
  \bibinfo{volume}{10}, \bibinfo{number}{2} (\bibinfo{year}{2012}),
  \bibinfo{pages}{71--75}.
\newblock


\bibitem[Balebako et~al\mbox{.}(2013)]%
        {Balbeko2013LittleBrothersWatchingYou}
\bibfield{author}{\bibinfo{person}{Rebecca Balebako}, \bibinfo{person}{Jaeyeon
  Jung}, \bibinfo{person}{Wei Lu}, \bibinfo{person}{Lorrie~Faith Cranor}, {and}
  \bibinfo{person}{Carolyn Nguyen}.} \bibinfo{year}{2013}\natexlab{}.
\newblock \showarticletitle{"Little Brothers Watching You": Raising Awareness
  of Data Leaks on Smartphones}. In \bibinfo{booktitle}{\emph{Proceedings of
  the Ninth Symposium on Usable Privacy and Security}} (Newcastle, United
  Kingdom) \emph{(\bibinfo{series}{SOUPS '13})}.
  \bibinfo{publisher}{Association for Computing Machinery},
  \bibinfo{address}{New York, NY, USA}, Article \bibinfo{articleno}{12},
  \bibinfo{numpages}{11}~pages.
\newblock
\showISBNx{9781450323192}


\bibitem[Breitinger et~al\mbox{.}(2019)]%
        {survey:breitinger:2019}
\bibfield{author}{\bibinfo{person}{Frank Breitinger}, \bibinfo{person}{Ryan
  Tully-Doyle}, {and} \bibinfo{person}{Courtney Hassenfeldt}.}
  \bibinfo{year}{2019}\natexlab{}.
\newblock \showarticletitle{A survey on smartphone user’s security choices,
  awareness and education}.
\newblock \bibinfo{journal}{\emph{Computers \& Security}}  \bibinfo{volume}{88}
  (\bibinfo{date}{10} \bibinfo{year}{2019}), \bibinfo{pages}{101647}.
\newblock


\bibitem[Campbell and Ahmed(2010)]%
        {campbellecosystems2010}
\bibfield{author}{\bibinfo{person}{P.~R.~J. Campbell} {and}
  \bibinfo{person}{Faheem Ahmed}.} \bibinfo{year}{2010}\natexlab{}.
\newblock \showarticletitle{A Three-Dimensional View of Software Ecosystems}.
  In \bibinfo{booktitle}{\emph{Proceedings of the Fourth European Conference on
  Software Architecture: Companion Volume}} (Copenhagen, Denmark)
  \emph{(\bibinfo{series}{ECSA '10})}. \bibinfo{publisher}{Association for
  Computing Machinery}, \bibinfo{address}{New York, NY, USA},
  \bibinfo{pages}{81–84}.
\newblock


\bibitem[Coen et~al\mbox{.}(2016)]%
        {Coen2016paradox}
\bibfield{author}{\bibinfo{person}{Rena Coen}, \bibinfo{person}{Jennifer King},
  {and} \bibinfo{person}{Richmond Wong}.} \bibinfo{year}{2016}\natexlab{}.
\newblock \showarticletitle{The Privacy Policy Paradox}. In
  \bibinfo{booktitle}{\emph{Twelfth Symposium on Usable Privacy and Security
  (SOUPS 2016)}}. \bibinfo{publisher}{USENIX Association},
  \bibinfo{address}{Denver, CO}, \bibinfo{numpages}{3}~pages.
\newblock


\bibitem[Competition and Markets~Authority(2021)]%
        {mobilecosystemsstudy}
\bibfield{author}{\bibinfo{person}{Competition} {and} \bibinfo{person}{GOV.UK
  Markets~Authority}.} \bibinfo{year}{2021}\natexlab{}.
\newblock \bibinfo{booktitle}{\emph{Mobile ecosystems Market Study}}.
\newblock
\urldef\tempurl%
\url{https://assets.publishing.service.gov.uk/government/uploads/system/uploads/attachment\_data/file/1048746/MobileEcosystems\_InterimReport.pdf}
\showURL{%
Retrieved June 7, 2022 from \tempurl}
\newblock
\shownote{Web link}.


\bibitem[Consolvo et~al\mbox{.}(2005)]%
        {Consolvo2005locationdisclosure}
\bibfield{author}{\bibinfo{person}{Sunny Consolvo}, \bibinfo{person}{Ian~E.
  Smith}, \bibinfo{person}{Tara Matthews}, \bibinfo{person}{Anthony LaMarca},
  \bibinfo{person}{Jason Tabert}, {and} \bibinfo{person}{Pauline Powledge}.}
  \bibinfo{year}{2005}\natexlab{}.
\newblock \showarticletitle{Location Disclosure to Social Relations: Why, When,
  \& What People Want to Share}. In \bibinfo{booktitle}{\emph{Proceedings of
  the SIGCHI Conference on Human Factors in Computing Systems}} (Portland,
  Oregon, USA) \emph{(\bibinfo{series}{CHI '05})}.
  \bibinfo{publisher}{Association for Computing Machinery},
  \bibinfo{address}{New York, NY, USA}, \bibinfo{pages}{81–90}.
\newblock


\bibitem[Conti and Sobiesk(2010)]%
        {conti:2010}
\bibfield{author}{\bibinfo{person}{Gregory Conti} {and} \bibinfo{person}{Edward
  Sobiesk}.} \bibinfo{year}{2010}\natexlab{}.
\newblock \showarticletitle{Malicious Interface Design: Exploiting the User}.
  In \bibinfo{booktitle}{\emph{Proceedings of the 19th International Conference
  on World Wide Web}} (Raleigh, North Carolina, USA)
  \emph{(\bibinfo{series}{WWW '10})}. \bibinfo{publisher}{Association for
  Computing Machinery}, \bibinfo{address}{New York, NY, USA},
  \bibinfo{pages}{271–280}.
\newblock
\showISBNx{9781605587998}


\bibitem[Coopamootoo et~al\mbox{.}(2022)]%
        {Ifeelinvadedannoyed:2022:usenix}
\bibfield{author}{\bibinfo{person}{Kovila~P.L. Coopamootoo},
  \bibinfo{person}{Maryam Mehrnezhad}, {and} \bibinfo{person}{Ehsan Toreini}.}
  \bibinfo{year}{2022}\natexlab{}.
\newblock \showarticletitle{"I feel invaded, annoyed, anxious and I may protect
  myself": Individuals{\textquoteright} Feelings about Online Tracking and
  their Protective Behaviour across Gender and Country}. In
  \bibinfo{booktitle}{\emph{31st USENIX Security Symposium (USENIX Security
  22)}}. \bibinfo{publisher}{USENIX Association}, \bibinfo{address}{Boston,
  MA}, \bibinfo{pages}{287--304}.
\newblock


\bibitem[Egelman et~al\mbox{.}(2013)]%
        {priceforthat:serge:2013}
\bibfield{author}{\bibinfo{person}{Serge Egelman}, \bibinfo{person}{Adrienne
  Felt}, {and} \bibinfo{person}{David Wagner}.}
  \bibinfo{year}{2013}\natexlab{}.
\newblock \showarticletitle{Choice Architecture and Smartphone Privacy: There's
  A Price for That}.
\newblock \bibinfo{journal}{\emph{The Economics of Information Security and
  Privacy}} (\bibinfo{date}{10} \bibinfo{year}{2013}),
  \bibinfo{pages}{211–236}.
\newblock


\bibitem[Felt et~al\mbox{.}(2011)]%
        {feltandroidpermissions2011}
\bibfield{author}{\bibinfo{person}{Adrienne~Porter Felt},
  \bibinfo{person}{Erika Chin}, \bibinfo{person}{Steve Hanna},
  \bibinfo{person}{Dawn Song}, {and} \bibinfo{person}{David Wagner}.}
  \bibinfo{year}{2011}\natexlab{}.
\newblock \showarticletitle{Android Permissions Demystified}
  \emph{(\bibinfo{series}{CCS '11})}. \bibinfo{publisher}{Association for
  Computing Machinery}, \bibinfo{address}{New York, NY, USA},
  \bibinfo{pages}{627–638}.
\newblock


\bibitem[Felt et~al\mbox{.}(2012a)]%
        {felt201299problems}
\bibfield{author}{\bibinfo{person}{Adrienne~Porter Felt},
  \bibinfo{person}{Serge Egelman}, {and} \bibinfo{person}{David Wagner}.}
  \bibinfo{year}{2012}\natexlab{a}.
\newblock \showarticletitle{I've Got 99 Problems, but Vibration Ain't One: A
  Survey of Smartphone Users' Concerns}. In
  \bibinfo{booktitle}{\emph{Proceedings of the Second ACM Workshop on Security
  and Privacy in Smartphones and Mobile Devices}} (Raleigh, North Carolina,
  USA) \emph{(\bibinfo{series}{SPSM '12})}. \bibinfo{publisher}{Association for
  Computing Machinery}, \bibinfo{address}{New York, NY, USA},
  \bibinfo{pages}{33–44}.
\newblock


\bibitem[Felt et~al\mbox{.}(2012b)]%
        {Felt2012androidpermissions}
\bibfield{author}{\bibinfo{person}{Adrienne~Porter Felt},
  \bibinfo{person}{Elizabeth Ha}, \bibinfo{person}{Serge Egelman},
  \bibinfo{person}{Ariel Haney}, \bibinfo{person}{Erika Chin}, {and}
  \bibinfo{person}{David Wagner}.} \bibinfo{year}{2012}\natexlab{b}.
\newblock \showarticletitle{Android Permissions: User Attention, Comprehension,
  and Behavior}. In \bibinfo{booktitle}{\emph{Proceedings of the Eighth
  Symposium on Usable Privacy and Security}} (Washington, D.C.)
  \emph{(\bibinfo{series}{SOUPS '12})}. \bibinfo{publisher}{Association for
  Computing Machinery}, \bibinfo{address}{New York, NY, USA}, Article
  \bibinfo{articleno}{3}, \bibinfo{numpages}{14}~pages.
\newblock


\bibitem[Frik et~al\mbox{.}(2022)]%
        {smartphoneprivacy:2022:Frik}
\bibfield{author}{\bibinfo{person}{Alisa Frik}, \bibinfo{person}{Juliann Kim},
  \bibinfo{person}{Joshua~Rafael Sanchez}, {and} \bibinfo{person}{Joanne Ma}.}
  \bibinfo{year}{2022}\natexlab{}.
\newblock \showarticletitle{Users’ Expectations About and Use of Smartphone
  Privacy and Security Settings}. In \bibinfo{booktitle}{\emph{CHI Conference
  on Human Factors in Computing Systems}} (New Orleans, LA, USA)
  \emph{(\bibinfo{series}{CHI '22})}. \bibinfo{publisher}{Association for
  Computing Machinery}, \bibinfo{address}{New York, NY, USA}, Article
  \bibinfo{articleno}{407}, \bibinfo{numpages}{24}~pages.
\newblock
\showISBNx{9781450391573}


\bibitem[Gamba et~al\mbox{.}(2020)]%
        {Gamba:2020:preinstalledapps}
\bibfield{author}{\bibinfo{person}{Julien Gamba}, \bibinfo{person}{Mohammed
  Rashed}, \bibinfo{person}{Abbas Razaghpanah}, \bibinfo{person}{Juan
  Tapiador}, {and} \bibinfo{person}{Narseo Vallina-Rodriguez}.}
  \bibinfo{year}{2020}\natexlab{}.
\newblock \showarticletitle{An Analysis of Pre-installed Android Software}. In
  \bibinfo{booktitle}{\emph{2020 IEEE Symposium on Security and Privacy (SP)}}.
  \bibinfo{pages}{1039--1055}.
\newblock


\bibitem[Google(2022)]%
        {googleecosystem}
\bibfield{author}{\bibinfo{person}{Google}.} \bibinfo{year}{2022}\natexlab{}.
\newblock \bibinfo{booktitle}{\emph{Google Products}}.
\newblock
\urldef\tempurl%
\url{https://about.google/products/}
\showURL{%
Retrieved February 18, 2022 from \tempurl}
\newblock
\shownote{Web link}.


\bibitem[Guest et~al\mbox{.}(2006)]%
        {howmanyinterviewsareenough:2006}
\bibfield{author}{\bibinfo{person}{Greg Guest}, \bibinfo{person}{Arwen Bunce},
  {and} \bibinfo{person}{Laura Johnson}.} \bibinfo{year}{2006}\natexlab{}.
\newblock \showarticletitle{How Many Interviews Are Enough?: An Experiment with
  Data Saturation and Variability}.
\newblock \bibinfo{journal}{\emph{Field Methods}} \bibinfo{volume}{18},
  \bibinfo{number}{1} (\bibinfo{year}{2006}), \bibinfo{pages}{59--82}.
\newblock


\bibitem[Hartmans(2021)]%
        {buisnessinsideronline2021}
\bibfield{author}{\bibinfo{person}{Avery Hartmans}.}
  \bibinfo{year}{2021}\natexlab{}.
\newblock \bibinfo{booktitle}{\emph{Apple said the Pro models of the iPhone 12
  sold the best last quarter as a record number of people upgraded their
  devices}}.
\newblock
\urldef\tempurl%
\url{https://www.businessinsider.com/apple-iphone-12-pro-pro-max-sold-best-in-q1-2021-1?r=US\&IR=T}
\showURL{%
Retrieved February 7, 2022 from \tempurl}
\newblock
\shownote{Web link}.


\bibitem[Huawei(2022)]%
        {huaweiecosystem}
\bibfield{author}{\bibinfo{person}{Huawei}.} \bibinfo{year}{2022}\natexlab{}.
\newblock \bibinfo{booktitle}{\emph{Huawei Ecosystem}}.
\newblock
\urldef\tempurl%
\url{https://consumer.huawei.com/en/community/details/Huawei-Ecosystem-Interconnectedness-of-my-devices/topicId_148614/}
\showURL{%
Retrieved February 18, 2022 from \tempurl}
\newblock
\shownote{Web link}.


\bibitem[Kang et~al\mbox{.}(2015)]%
        {kang2015dataeverywhere}
\bibfield{author}{\bibinfo{person}{Ruogu Kang}, \bibinfo{person}{Laura
  Dabbish}, \bibinfo{person}{Nathaniel Fruchter}, {and} \bibinfo{person}{Sara
  Kiesler}.} \bibinfo{year}{2015}\natexlab{}.
\newblock \showarticletitle{"My Data Just Goes Everywhere": User Mental Models
  of the Internet and Implications for Privacy and Security}. In
  \bibinfo{booktitle}{\emph{Proceedings of the Eleventh USENIX Conference on
  Usable Privacy and Security}} (Ottawa, Canada) \emph{(\bibinfo{series}{SOUPS
  '15})}. \bibinfo{publisher}{USENIX Association}, \bibinfo{address}{USA},
  \bibinfo{pages}{39–52}.
\newblock
\showISBNx{9781931971249}


\bibitem[Kelley et~al\mbox{.}(2009)]%
        {Kelley2009Nutritionlabel}
\bibfield{author}{\bibinfo{person}{Patrick~Gage Kelley},
  \bibinfo{person}{Joanna Bresee}, \bibinfo{person}{Lorrie~Faith Cranor}, {and}
  \bibinfo{person}{Robert~W. Reeder}.} \bibinfo{year}{2009}\natexlab{}.
\newblock \showarticletitle{A "Nutrition Label" for Privacy}. In
  \bibinfo{booktitle}{\emph{Proceedings of the 5th Symposium on Usable Privacy
  and Security}} (Mountain View, California, USA) \emph{(\bibinfo{series}{SOUPS
  '09})}. \bibinfo{publisher}{Association for Computing Machinery},
  \bibinfo{address}{New York, NY, USA}, Article \bibinfo{articleno}{4},
  \bibinfo{numpages}{12}~pages.
\newblock
\showISBNx{9781605587363}


\bibitem[Kelly et~al\mbox{.}(2012)]%
        {Kelley2012conundrumofpermissions}
\bibfield{author}{\bibinfo{person}{Patrick Kelly}, \bibinfo{person}{Sunny
  Consolvo}, \bibinfo{person}{Lorrie Cranor}, \bibinfo{person}{Jaeyeon Jung},
  \bibinfo{person}{Norman Sadeh}, {and} \bibinfo{person}{David Wetherall}.}
  \bibinfo{year}{2012}\natexlab{}.
\newblock \showarticletitle{A Conundrum of Permissions: Installing Applications
  on an Android Smartphone}.
\newblock \bibinfo{journal}{\emph{Proceedings of USEC 2012}}
  \bibinfo{volume}{7398}, \bibinfo{pages}{68--79}.
\newblock


\bibitem[Liccardi et~al\mbox{.}(2014)]%
        {notechnicalunderstandingrequired}
\bibfield{author}{\bibinfo{person}{Ilaria Liccardi}, \bibinfo{person}{Joseph
  Pato}, \bibinfo{person}{Daniel~J. Weitzner}, \bibinfo{person}{Hal Abelson},
  {and} \bibinfo{person}{David De~Roure}.} \bibinfo{year}{2014}\natexlab{}.
\newblock \showarticletitle{No Technical Understanding Required: Helping Users
  Make Informed Choices about Access to Their Personal Data}. In
  \bibinfo{booktitle}{\emph{Proceedings of the 11th International Conference on
  Mobile and Ubiquitous Systems: Computing, Networking and Services}} (London,
  United Kingdom) \emph{(\bibinfo{series}{MOBIQUITOUS '14})}.
  \bibinfo{publisher}{ICST (Institute for Computer Sciences, Social-Informatics
  and Telecommunications Engineering)}, \bibinfo{address}{Brussels, BEL},
  \bibinfo{pages}{140–150}.
\newblock
\showISBNx{9781631900396}


\bibitem[Liu et~al\mbox{.}(2014)]%
        {privacyprofiles:liu}
\bibfield{author}{\bibinfo{person}{Bin Liu}, \bibinfo{person}{Jialiu Lin},
  {and} \bibinfo{person}{Norman Sadeh}.} \bibinfo{year}{2014}\natexlab{}.
\newblock \showarticletitle{Reconciling Mobile App Privacy and Usability on
  Smartphones: Could User Privacy Profiles Help?}. In
  \bibinfo{booktitle}{\emph{Proceedings of the 23rd International Conference on
  World Wide Web}} (Seoul, Korea) \emph{(\bibinfo{series}{WWW '14})}.
  \bibinfo{publisher}{Association for Computing Machinery},
  \bibinfo{address}{New York, NY, USA}, \bibinfo{pages}{201–212}.
\newblock
\showISBNx{9781450327442}


\bibitem[Lutaaya(2018)]%
        {rethinkingapppermissions}
\bibfield{author}{\bibinfo{person}{Michael Lutaaya}.}
  \bibinfo{year}{2018}\natexlab{}.
\newblock \showarticletitle{Rethinking App Permissions on IOS}. In
  \bibinfo{booktitle}{\emph{Extended Abstracts of the 2018 CHI Conference on
  Human Factors in Computing Systems}} (Montreal QC, Canada)
  \emph{(\bibinfo{series}{CHI EA '18})}. \bibinfo{publisher}{Association for
  Computing Machinery}, \bibinfo{address}{New York, NY, USA},
  \bibinfo{pages}{1–6}.
\newblock
\showISBNx{9781450356213}


\bibitem[Malkin et~al\mbox{.}(2019)]%
        {Malkin:2019:smartspeakers}
\bibfield{author}{\bibinfo{person}{Nathan Malkin}, \bibinfo{person}{Joe
  Deatrick}, \bibinfo{person}{Allen Tong}, \bibinfo{person}{Primal Wijesekera},
  \bibinfo{person}{Serge Egelman}, {and} \bibinfo{person}{David Wagner}.}
  \bibinfo{year}{2019}\natexlab{}.
\newblock \showarticletitle{Privacy Attitudes of Smart Speaker Users}.
\newblock \bibinfo{journal}{\emph{Proceedings on Privacy Enhancing
  Technologies}}  \bibinfo{volume}{2019} (\bibinfo{date}{10}
  \bibinfo{year}{2019}), \bibinfo{pages}{250--271}.
\newblock


\bibitem[McDonald and Cranor(2010)]%
        {McDonald2010BeliefsAB:2010}
\bibfield{author}{\bibinfo{person}{Aleecia McDonald} {and}
  \bibinfo{person}{Lorrie~Faith Cranor}.} \bibinfo{year}{2010}\natexlab{}.
\newblock \showarticletitle{Beliefs and Behaviors: Internet Users'
  Understanding of Behavioral Advertising}. \bibinfo{publisher}{TPRC},
  \bibinfo{pages}{1--31}.
\newblock


\bibitem[Palmerino(2018)]%
        {improvingpermissions:palmerino}
\bibfield{author}{\bibinfo{person}{Jeffrey Palmerino}.}
  \bibinfo{year}{2018}\natexlab{}.
\newblock \showarticletitle{Improving Android Permissions Models for Increased
  User Awareness and Security}. In \bibinfo{booktitle}{\emph{Proceedings of the
  5th International Conference on Mobile Software Engineering and Systems}}
  (Gothenburg, Sweden) \emph{(\bibinfo{series}{MOBILESoft '18})}.
  \bibinfo{publisher}{Association for Computing Machinery},
  \bibinfo{address}{New York, NY, USA}, \bibinfo{pages}{41–42}.
\newblock


\bibitem[Ramokapane et~al\mbox{.}(2019)]%
        {Ramokapane2019}
\bibfield{author}{\bibinfo{person}{Kopo~M. Ramokapane},
  \bibinfo{person}{Anthony~C. Mazeli}, {and} \bibinfo{person}{Awais Rashid}.}
  \bibinfo{year}{2019}\natexlab{}.
\newblock \showarticletitle{Skip, Skip, Skip, Accept: A Study on the Usability
  of Smartphone Manufacturer Provided Default Features and User Privacy}.
\newblock \bibinfo{journal}{\emph{Proceedings on Privacy Enhancing
  Technologies}} \bibinfo{volume}{2019}, \bibinfo{number}{2}
  (\bibinfo{date}{April} \bibinfo{year}{2019}), \bibinfo{pages}{209 -- 227}.
\newblock


\bibitem[Redmiles et~al\mbox{.}(2018)]%
        {redmiles:2018:biases}
\bibfield{author}{\bibinfo{person}{Elissa~M. Redmiles}, \bibinfo{person}{Ziyun
  Zhu}, \bibinfo{person}{Sean Kross}, \bibinfo{person}{Dhruv Kuchhal},
  \bibinfo{person}{Tudor Dumitras}, {and} \bibinfo{person}{Michelle~L.
  Mazurek}.} \bibinfo{year}{2018}\natexlab{}.
\newblock \showarticletitle{Asking for a Friend: Evaluating Response Biases in
  Security User Studies}. In \bibinfo{booktitle}{\emph{Proceedings of the 2018
  ACM SIGSAC Conference on Computer and Communications Security}} (Toronto,
  Canada) \emph{(\bibinfo{series}{CCS '18})}. \bibinfo{publisher}{Association
  for Computing Machinery}, \bibinfo{address}{New York, NY, USA},
  \bibinfo{pages}{1238–1255}.
\newblock
\showISBNx{9781450356930}


\bibitem[Sambasivan et~al\mbox{.}(2018)]%
        {Sambasivan:2018:richwomen}
\bibfield{author}{\bibinfo{person}{Nithya Sambasivan}, \bibinfo{person}{Garen
  Checkley}, \bibinfo{person}{Amna Batool}, \bibinfo{person}{Nova Ahmed},
  \bibinfo{person}{David Nemer}, \bibinfo{person}{Laura~Sanely
  Gayt\'{a}n-Lugo}, \bibinfo{person}{Tara Matthews}, \bibinfo{person}{Sunny
  Consolvo}, {and} \bibinfo{person}{Elizabeth Churchil}.}
  \bibinfo{year}{2018}\natexlab{}.
\newblock \showarticletitle{"Privacy is Not for Me, It's for Those Rich Women":
  Performative Privacy Practices on Mobile Phones by Women in South Asia}. In
  \bibinfo{booktitle}{\emph{Proceedings of the Fourteenth USENIX Conference on
  Usable Privacy and Security}} (Baltimore, MD, USA)
  \emph{(\bibinfo{series}{SOUPS '18})}. \bibinfo{publisher}{USENIX
  Association}, \bibinfo{address}{USA}, \bibinfo{pages}{127–142}.
\newblock
\showISBNx{9781931971454}


\bibitem[Shen et~al\mbox{.}(2021)]%
        {Shen2021}
\bibfield{author}{\bibinfo{person}{Bingyu~We Shen}, \bibinfo{person}{Lili
  Xiang}, \bibinfo{person}{Chengcheng Wu}, \bibinfo{person}{Yudong Shen},
  \bibinfo{person}{Mingyao Zhou}, \bibinfo{person}{Yuanyuan Jin}, {and}
  \bibinfo{person}{Xinxin}.} \bibinfo{year}{2021}\natexlab{}.
\newblock \showarticletitle{Can Systems Explain Permissions Better?
  Understanding Users Misperceptions under Smartphone Runtime Permission
  Model}. In \bibinfo{booktitle}{\emph{30th {USENIX} Security Symposium
  ({USENIX} Security 21)}}. \bibinfo{publisher}{{USENIX} Association},
  \bibinfo{address}{Vancouver, B.C.}
\newblock


\bibitem[Shklovski et~al\mbox{.}(2014)]%
        {Leakinessandcreepiness:irina}
\bibfield{author}{\bibinfo{person}{Irina Shklovski}, \bibinfo{person}{Scott~D.
  Mainwaring}, \bibinfo{person}{Halla~Hrund Sk\'{u}lad\'{o}ttir}, {and}
  \bibinfo{person}{H\"{o}skuldur Borgthorsson}.}
  \bibinfo{year}{2014}\natexlab{}.
\newblock \showarticletitle{Leakiness and Creepiness in App Space: Perceptions
  of Privacy and Mobile App Use}. In \bibinfo{booktitle}{\emph{Proceedings of
  the SIGCHI Conference on Human Factors in Computing Systems}} (Toronto,
  Canada) \emph{(\bibinfo{series}{CHI '14})}. \bibinfo{publisher}{Association
  for Computing Machinery}, \bibinfo{address}{New York, NY, USA},
  \bibinfo{pages}{2347–2356}.
\newblock
\showISBNx{9781450324731}


\bibitem[Smullen et~al\mbox{.}(2020)]%
        {smullen:2020}
\bibfield{author}{\bibinfo{person}{Daniel Smullen}, \bibinfo{person}{Yuanyuan
  Feng}, \bibinfo{person}{Shikun~Aerin Zhang}, {and} \bibinfo{person}{Norman
  Sadeh}.} \bibinfo{year}{2020}\natexlab{}.
\newblock \showarticletitle{The Best of Both Worlds: Mitigating Trade-offs
  Between Accuracy and User Burden in Capturing Mobile App Privacy
  Preferences}.
\newblock \bibinfo{journal}{\emph{Proceedings on Privacy Enhancing
  Technologies}} \bibinfo{volume}{2020}, \bibinfo{number}{1}
  (\bibinfo{year}{2020}), \bibinfo{pages}{195–215}.
\newblock


\bibitem[Statista(2021)]%
        {surveyagegroup2019}
\bibfield{author}{\bibinfo{person}{Statista}.} \bibinfo{year}{2021}\natexlab{}.
\newblock \bibinfo{booktitle}{\emph{Smartphone OS in 2019, by age group}}.
\newblock
\urldef\tempurl%
\url{https://www.statista.com/statistics/1133193/smartphone-os-by-age/}
\showURL{%
Retrieved February 8, 2022 from \tempurl}
\newblock
\shownote{Web link}.


\bibitem[Tan et~al\mbox{.}(2014)]%
        {joshuaexplanationforpermission2014}
\bibfield{author}{\bibinfo{person}{Joshua Tan}, \bibinfo{person}{Khanh Nguyen},
  \bibinfo{person}{Michael Theodorides}, \bibinfo{person}{Heidi
  Negr\'{o}n-Arroyo}, \bibinfo{person}{Christopher Thompson},
  \bibinfo{person}{Serge Egelman}, {and} \bibinfo{person}{David Wagner}.}
  \bibinfo{year}{2014}\natexlab{}.
\newblock \showarticletitle{The Effect of Developer-Specified Explanations for
  Permission Requests on Smartphone User Behavior}. In
  \bibinfo{booktitle}{\emph{Proceedings of the SIGCHI Conference on Human
  Factors in Computing Systems}} (Toronto, Ontario, Canada)
  \emph{(\bibinfo{series}{CHI '14})}. \bibinfo{publisher}{Association for
  Computing Machinery}, \bibinfo{address}{New York, NY, USA},
  \bibinfo{pages}{91–100}.
\newblock


\bibitem[Thompson et~al\mbox{.}(2013)]%
        {forgivness:thompson}
\bibfield{author}{\bibinfo{person}{Christopher Thompson},
  \bibinfo{person}{Maritza Johnson}, \bibinfo{person}{Serge Egelman},
  \bibinfo{person}{David Wagner}, {and} \bibinfo{person}{Jennifer King}.}
  \bibinfo{year}{2013}\natexlab{}.
\newblock \showarticletitle{When It's Better to Ask Forgiveness than Get
  Permission: Attribution Mechanisms for Smartphone Resources}. In
  \bibinfo{booktitle}{\emph{Proceedings of the Ninth Symposium on Usable
  Privacy and Security}} (Newcastle, UK) \emph{(\bibinfo{series}{SOUPS '13})}.
  \bibinfo{publisher}{Association for Computing Machinery},
  \bibinfo{address}{New York, NY, USA}, Article \bibinfo{articleno}{1},
  \bibinfo{numpages}{14}~pages.
\newblock
\showISBNx{9781450323192}


\bibitem[Ur et~al\mbox{.}(2012)]%
        {smartusefulcreepy:2012:blase}
\bibfield{author}{\bibinfo{person}{Blase Ur}, \bibinfo{person}{Pedro~Giovanni
  Leon}, \bibinfo{person}{Lorrie~Faith Cranor}, \bibinfo{person}{Richard Shay},
  {and} \bibinfo{person}{Yang Wang}.} \bibinfo{year}{2012}\natexlab{}.
\newblock \showarticletitle{Smart, Useful, Scary, Creepy: Perceptions of Online
  Behavioral Advertising}. In \bibinfo{booktitle}{\emph{Proceedings of the
  Eighth Symposium on Usable Privacy and Security}} (Washington, D.C.)
  \emph{(\bibinfo{series}{SOUPS '12})}. \bibinfo{publisher}{Association for
  Computing Machinery}, \bibinfo{address}{New York, NY, USA}, Article
  \bibinfo{articleno}{4}, \bibinfo{numpages}{15}~pages.
\newblock
\showISBNx{9781450315326}


\bibitem[Van~Kleek et~al\mbox{.}(2017)]%
        {Van2017BetterTheDevilYouKnow}
\bibfield{author}{\bibinfo{person}{Max Van~Kleek}, \bibinfo{person}{Ilaria
  Liccardi}, \bibinfo{person}{Reuben Binns}, \bibinfo{person}{Jun Zhao},
  \bibinfo{person}{Daniel~J. Weitzner}, {and} \bibinfo{person}{Nigel
  Shadbolt}.} \bibinfo{year}{2017}\natexlab{}.
\newblock \showarticletitle{Better the Devil You Know: Exposing the Data
  Sharing Practices of Smartphone Apps}. In
  \bibinfo{booktitle}{\emph{Proceedings of the 2017 CHI Conference on Human
  Factors in Computing Systems}} (Denver, Colorado, USA)
  \emph{(\bibinfo{series}{CHI '17})}. \bibinfo{publisher}{Association for
  Computing Machinery}, \bibinfo{address}{New York, NY, USA},
  \bibinfo{pages}{5208–5220}.
\newblock
\showISBNx{9781450346559}


\bibitem[Wijesekera et~al\mbox{.}(2018)]%
        {Wijesekera2018PrivacyDecisionsPredicition}
\bibfield{author}{\bibinfo{person}{Primal Wijesekera}, \bibinfo{person}{Joel
  Reardon}, \bibinfo{person}{Irwin Reyes}, \bibinfo{person}{Lynn Tsai},
  \bibinfo{person}{Jung-Wei Chen}, \bibinfo{person}{Nathan Good},
  \bibinfo{person}{David Wagner}, \bibinfo{person}{Konstantin Beznosov}, {and}
  \bibinfo{person}{Serge Egelman}.} \bibinfo{year}{2018}\natexlab{}.
\newblock \showarticletitle{Contextualizing Privacy Decisions for Better
  Prediction (and Protection)}. In \bibinfo{booktitle}{\emph{Proceedings of the
  2018 CHI Conference on Human Factors in Computing Systems}} (Montreal QC,
  Canada) \emph{(\bibinfo{series}{CHI '18})}. \bibinfo{publisher}{Association
  for Computing Machinery}, \bibinfo{address}{New York, NY, USA},
  \bibinfo{pages}{1–13}.
\newblock


\bibitem[Wikipedia(2021)]%
        {iMessage:Wiki}
\bibfield{author}{\bibinfo{person}{Wikipedia}.}
  \bibinfo{year}{2021}\natexlab{}.
\newblock \bibinfo{booktitle}{\emph{iMessage}}.
\newblock
\urldef\tempurl%
\url{https://en.wikipedia.org/wiki/IMessage}
\showURL{%
Retrieved April 12, 2021 from \tempurl}
\newblock
\shownote{Web link}.


\end{thebibliography}

\newpage
\appendix

\section{Appendix}

%\subsection{Appendix A: Online Survey}
%\label{appendixA}
%\input{survey}

%\subsection{Appendix B: Interview Guide}

%\label{appendixB}
%\input{InterviewGuide}

%\begin{figure}[h]
%  \centering
%  \includegraphics[width=\linewidth]{usenix2022_soups_latex-template/Diagrams and Figures/pre-survey(1).png}
%  \caption{Screenshot 1 of online pre-survey used for pre-screening participants}
%\end{figure}

%\begin{figure}[h]
%  \centering
%  \includegraphics[width=\linewidth]{usenix2022_soups_latex-template/Diagrams and Figures/pre-survey(2).png}
%  \caption{Screenshot 2 of online pre-survey used for pre-screening participants}
%\end{figure}

\subsection{Privacy Controls}
\label{appendixC}

\begin{table*}[htbp]
  \centering
  \caption{Privacy Configurations for each default app: Safari, Siri, iMessage and Facetime. N steps refer to the minimum number of steps or configuration paths suggested by this work to disable a default app (See Tables \ref{tab:privacycontrols1:app} and \ref{tab:privacycontrols2:app} for paths to disable settings of a default app). Last column suggests whether users' personal data may leave their devices. Yes/No: Apple mentions this information in their Privacy Policy documents, Information not provided by vendor: there is currently no information on Apple's Privacy Policy or resources if this data category leaves users' device or not. Not Applicable: no known personal data involved.}
    \begin{tabular}{llll}
    \toprule
    \textbf{App} & \textbf{N Steps} & \textbf{Privacy Configurations} & \textbf{May Transfer to Cloud or Vendor's Servers} \\
    \midrule
    Safari & N>12  & IP Address & Yes \\
          &       & Fradulent Website Warning & Not Applicable \\
          &       & Privacy Preserving Ad Measurement & Yes \\
          &       & Check for Apple Pay & Yes \\
          &       & Private Browsing & No \\
          &       & Web Page Translation & Translation locally, other data may leave device \\
          &       & Web Extensions & Yes \\
          &       & iCloud Syncing & Yes \\
          &       & Search Engine Suggestions & Yes \\
          &       & Preload Top Hit in Safari & Information not provided by vendor \\
          &       & Sending Information to Apple & Yes \\
          &       & History and Website Data & Yes \\
    \midrule
    Siri  & N>9   & Ask Siri & Yes \\
          &       & Integrated apps & Yes \\
          &       & Dictation & Yes \\
          &       & Siri and Dictation & Yes \\
          &       & Siri Personalisation & Yes \\
          &       & iCloud Syncing & Yes \\
          &       & Transcription & Yes \\
          &       & Location Services & Yes \\
          &       & Request History & Yes \\
    \midrule
    iMessage & N>5   & Messages in iCloud & Yes \\
          &       & Delete Messages & Yes \\
          &       & iCloud Backup & Yes \\
          &       & Shared with You & Yes \\
          &       & Shared with Apps & Yes \\
    \midrule
    Facetime & N>7   & Enable Facetime & Actual calls No, Otherwise Yes \\
          &       & Caller ID & Yes \\
          &       & SharePlay & Information not provided by vendor \\
          &       & Speaking & Information not provided by vendor \\
          &       & FaceTime Live Photos & Information not provided by vendor \\
          &       & Eye Contact & Information not provided by vendor \\
          &       & Blocked Contacts & Information not provided by vendor \\
    \bottomrule
    \end{tabular}%
  \label{tab:datacloud:appendix1}%
\end{table*}%

\begin{table*}[htbp]
  \centering
  \caption{Privacy Configurations for each default app: Touch ID, Location, Find My. N steps refer to the minimum number of steps or configuration paths suggested by this work to disable a default app (See Tables \ref{tab:privacycontrols1:app} and \ref{tab:privacycontrols2:app} for paths to disable settings of a default app). Last column suggests whether users' personal data may leave their devices. Yes/No: Apple mentions this information in their Privacy Policy documents, Information not provided by vendor: there is currently no information on Apple's Privacy Policy or resources if this data category leaves users' device or not. Not Applicable: no known personal data involved.}
    \begin{tabular}{llll}
    \toprule
    \textbf{App} & \textbf{N Steps} & \textbf{Privacy Configurations} & \textbf{May Transfer to Cloud or Vendor's Servers} \\
    \midrule
    Touch ID & N>9   & Passcode & No \\
          &       & Add fingerprint (1,2,..N=5 max) & No \\
          &       & Delete fingerprint & No \\
          &       & Device Unlock & No \\
          &       & iTunes and App Store & No \\
          &       & Wallet and Apple Pay & No \\
          &       & Password AutoFill & No \\
          &       & Voice Dial & No \\
          &       & Allow Access to Apps & Information not provided by vendor \\
    \midrule
    Location & N>6   & Enable Location & Yes \\
          &       & Tracking & Yes \\
          &       & Allow Apps to Track & Yes \\
          &       & Analytics and Improvements & Yes \\
          &       & Apple Advertising & Yes \\
          &       & App Privacy Report & Yes \\
    \midrule
    Find My & N>4   & Enable Find My & Yes \\
          &       & My Location & Yes \\
          &       & Share My Location & Yes \\
          &       & Family & Yes \\
    \bottomrule
    \end{tabular}%
  \label{tab:datacloud:appendix2}%
\end{table*}%

%% Privacy configurations
\begin{table*}[]
  \centering
  \caption{Privacy configurations of default apps: Safari, Siri and iMessage and paths to disable these settings. Paths marked by (*) are provided in Apple's Privacy Policy. The unmarked paths are the suggestions of this work to disable the settings. The unmarked paths are not provided for users on official documents.}
    \begin{tabular}{rll}
    \midrule
    \multicolumn{1}{p{4.085em}}{\textbf{Default App}} & \multicolumn{1}{c}{\textbf{Privacy Configurations}} & \multicolumn{1}{c}{\textbf{Path to Disable}} \\
    \midrule
    \multicolumn{1}{p{4.085em}}{Safari} & Hide IP Address* & Settings > Safari > Hide IP Address \\
\cmidrule{2-3}          & Fradulent Website Warning* & Settings > Safari > Fradulent Website Warning \\
\cmidrule{2-3}          & Privacy Preserving Ad Measurement* & Settings > Safari > Privacy Preserving Ad Measurement \\
\cmidrule{2-3}          & Check for Apple Pay* & Settings > Safari > Privacy and Security \\
\cmidrule{2-3}          & Private Browsing & Safari > Tabs > Private \\
\cmidrule{2-3}          & Web Page Translation & Settings > Safari > Advanced > Website data  \\
          &       & > Search "translate" or "Google" \\
\cmidrule{2-3}          & Web Extensions* & Settings > Safari > Extensions \\
\cmidrule{2-3}          & iCloud Syncing* & Settings > [your name] > iCloud > Safari \\
\cmidrule{2-3}          & Search Engine Suggestions & Settings > Safari > Search > Safari Suggestions \\
\cmidrule{2-3}          & Preload Top Hit in Safari & Settings > Safari > Search > Preload Top Hit \\
\cmidrule{2-3}          & Siri Suggestions* & Settings > Safari > Safari Suggestions \\
\cmidrule{2-3}          & Clear History and Website Data & Settings > Safari > Clear History and Website Data \\
    \midrule
    \multicolumn{1}{p{4.085em}}{Siri} & Ask Siri* & \multicolumn{1}{p{25.665em}}{Settings > Siri and Search > Listen for "Hey Siri" } \\
          &       & Settings > Accessibility > Always Listen for "Hey Siri" \\
\cmidrule{2-3}          & Integrated apps & Settings > Siri and Search > [app name] > Use with Ask Siri \\
\cmidrule{2-3}          & Dictation & Settings > General > Keyboard > Enable Dictation \\
\cmidrule{2-3}          & Suggestions on Lock Screen* & Settings > Siri and Search > Turn off suggestions on Lock Screen \\
\cmidrule{2-3}          & Location Services for Siri Suggestions* & Settings > Privacy > Location Services > \\
         &  & System Services > Location-Based Suggestions \\
\cmidrule{2-3}          & Siri and Dictation* & Settings > Screen Time > Content and Privacy Restrictions  \\
          &       & > Allowed apps \\
\cmidrule{2-3}          & Siri Personalisation* & Settings > [your name] > iCloud > Siri \\
\cmidrule{2-3}          & iCloud Syncing* & Settings > [your name] > iCloud > Siri \\
\cmidrule{2-3}          & Transcription* & Settings > Privacy > Speech Recognition \\
\cmidrule{2-3}          & Location Services & Settings > Privacy > Location Privacy \\
\cmidrule{2-3}          & Request History & Settings > Siri and Search > Siri and Dictation History \\
    \midrule
    \multicolumn{1}{p{4.085em}}{iMessage} & Messages in iCloud* & Settings > Messages > iMessage \\
\cmidrule{2-3}          & Delete Messages & Settings > Messages > iMessage > Message History \\
\cmidrule{2-3}          & iCloud Backup & Settings > Messages > iMessage > Enable Messages in iCloud \\
\cmidrule{2-3}          & Shared with You* & Settings > Messages > Shared with You \\
\cmidrule{2-3}          & Shared with Apps & Settings > Messages > Shared with You > Apps \\
    \bottomrule
    \end{tabular}%
  \label{tab:privacycontrols1:app}%
\end{table*}%

%%%%%%%%%%%%%%%%%%%%%%%%%%%%%%%%%%%%%%%%%%%%%%%%%%%%%%%%%%%%%%%%%%%%%%%%%%%%%%%%

\begin{table*}[htbp]
  \centering
  \caption{Privacy configurations of default apps: Facetime, Family Sharing, Touch ID, Location and Find My and paths to disable these settings. Paths marked by (*) are provided in Apple's Privacy Policy. The unmarked paths are the suggestions of this work to disable the settings. The unmarked paths are not provided for users on official documents.}
    \begin{tabular}{rll}
    \toprule
    \multicolumn{1}{p{4.085em}}{\textbf{Default App}} & \multicolumn{1}{c}{\textbf{Privacy Configurations}} & \multicolumn{1}{c}{\textbf{Path to Disable}} \\
    \midrule
    \multicolumn{1}{p{4.085em}}{Facetime} & Enable Facetime* & Settings > FaceTime > Toggle FaceTime \\
\cmidrule{2-3}          & Caller ID & Settings > FaceTime > Caller ID \\
\cmidrule{2-3}          & SharePlay & Settings > FaceTime > SharePlay \\
\cmidrule{2-3}          & Speaking & Settings > FaceTime > Automatic Prominence > Speaking \\
\cmidrule{2-3}          & FaceTime Live Photos & Settings > FaceTime > FaceTime Live Photos \\
\cmidrule{2-3}          & Eye Contact & Settings > FaceTime > Eye Contact \\
\cmidrule{2-3}          & Blocked Contacts & Settings > FaceTime > Calls > Blocked Contacts \\
    \midrule
    \multicolumn{1}{p{4.085em}}{Family} & Family Setup & Settings > Apple ID > Family \\
\cmidrule{2-3}    \multicolumn{1}{p{4.085em}}{Sharing} & Apple Subscription & Settings > Apple ID > Family > Apple Subscriptions \\
\cmidrule{2-3}          & Ask to Buy & Settings > Apple ID > Family > App Store Subscriptions \\
\cmidrule{2-3}          & Screen Time & Settings > Apple ID > Family > Screen Time \\
\cmidrule{2-3}          & Purchase Sharing & Settings > Apple ID > Family > More to Share > Purchase Sharing \\
\cmidrule{2-3}          & iCloud+ & Settings > Apple ID > Family > More to Share > iCloud+ \\
    \midrule
    \multicolumn{1}{p{4.085em}}{Touch ID} & Passcode & Settings > Touch ID and Passcode > Turn Passcode Off \\
\cmidrule{2-3}          & Add fingerprint (1,2,..N=5 max) & Settings > Touch ID and Passcode > Fingerprints \\
\cmidrule{2-3}          & Delete fingerprint & Settings > Touch ID and Passcode > Fingerprints \\
\cmidrule{2-3}          & Device Unlock & Settings > Touch ID and Passcode > Use Touch ID For \\
\cmidrule{2-3}          & iTunes and App Store & Settings > Touch ID and Passcode > Use Touch ID For \\
\cmidrule{2-3}          & Wallet and Apple Pay & Settings > Touch ID and Passcode > Use Touch ID For \\
\cmidrule{2-3}          & Password AutoFill & Settings > Touch ID and Passcode > Use Touch ID For \\
\cmidrule{2-3}          & Voice Dial & Settings > Touch ID and Passcode > Voice Dial \\
\cmidrule{2-3}          & Allow Access to Apps & Settings > Touch ID and Passcode > Allow Access When Locked \\
    \midrule
    \multicolumn{1}{p{4.085em}}{Location} & Enable Location* & Settings > Privacy > Location Services \\
\cmidrule{2-3}          & Tracking & Settings > Privacy > Location Services > Location Alerts \\
\cmidrule{2-3}          & Allow Apps to Track & Settings > Privacy > Tracking > Allow Apps to Request Track \\
\cmidrule{2-3}          & Analytics and Improvements & Settings > Privacy > Analytics and Improvements \\
\cmidrule{2-3}          & Apple Advertising & Settings > Privacy > Apple Advertising \\
\cmidrule{2-3}          & App Privacy Report & Settings > Privacy > App Privacy Report \\
    \midrule
    \multicolumn{1}{p{4.085em}}{Find My} & Enable Find My* & Settings > Apple ID > Find My \\
\cmidrule{2-3}          & My Location & Settings > Apple ID > Find My > My Location \\
\cmidrule{2-3}          & Share My Location & Settings > Apple ID > Find My > Share My Location \\
\cmidrule{2-3}          & Family & Settings > Apple ID > Find My > Family \\
    \bottomrule
    \end{tabular}%
  \label{tab:privacycontrols2:app}%
\end{table*}%

%%%%%%%%%%%%%%%%%%%%%%%%%%%%%%%%%%%%%%%%%%%%%%%%%%%%%%%%%%%%%%%%%%%%%%%%%%%%%%%%

%%  LocalWords:  endnotes includegraphics fread ptr nobj noindent
%%  LocalWords:  pdflatex acks

\newpage 
\onecolumn
  \subsection{Participants details}
  \label{AppendixE}
  % Table generated by Excel2LaTeX from sheet 'Participant list'
\begin{table}[h]
  \centering
  \caption{Results of the screening survey show participants' occupations, iPhone Model, MacBook Model and default apps used. All iPhones used in this expirement run iOS 14.0+ and all Macbooks run macOS X 10.15+. Key for the default apps (last column): 1-TouchID, 2-FindMy, 3-Siri, 4-Safari, 5-Location Services, 6-Family Sharing, 7-iMessage, 8-Facetime.}
    \begin{tabular}{clllc}
    \toprule
    P\#   & \multicolumn{1}{c}{Occupation} & \multicolumn{1}{c}{iPhone Model} & \multicolumn{1}{c}{Macbook Model} & Default Apps \\
    \midrule
    P01   & PhD candidate in Computer Science & iPhone 6S & MacBook Pro: 2012+ & 1,4,7,8 \\
    P02   & Software Developer & iPhone X & MacBook: 2015+ & 1- 5,7,8 \\
    P03   & Industrial Engineering Researcher & iPhone SE (2016) & MacBook Pro: 2012+ & 4,5,7,8 \\
    P04   & GIS Specialist / MSc Student in Forestry &  iPhone 7 & MacBook Pro: 2012+ & 1-5,7,8 \\
    P05   & Educational Sciences Student & iPhone SE (2020) & MacBook Air: 2012+ & 1,2,4,5,8 \\
    P06   & UI/UX Designer, Front-End Web Developer & iPhone 11, Pro, Pro Max & MacBook Pro: 2012+ & 2-8 \\
    P07   & Hospitality Management Student &  iPhone 7 & MacBook: 2015+ & 1-4, 6-8 \\
    P08   & Architect &  iPhone 7 & MacBook Pro: 2012+ & 1-5, 7,8 \\
    P09   & Economist & iPhone SE (2016) & MacBook: 2015+ & 1 - 4, 7 \\
    P10   & Assistant & iPhone 11, Pro, Pro Max & MacBook: 2015+ & 1-5, 7,8 \\
    P11   & Graphic Designer & iPhone XR & MacBook Air: 2012+ & 1,2,4,5,7 \\
    P12   & Unemployed & iPhone 11, Pro, Pro Max & MacBook Air: 2012+ & 2 - 5,  \\
    P13   & Program Manager & iPhone 11, Pro, Pro Max & MacBook Pro: 2012+ & 1,2,4-8 \\
    P14   & Laboratory Technician (sculptor) & iPhone 11, Pro, Pro Max & MacBook Pro: 2012+ & 1-5, 7,8 \\
    P15   & Document Controller & iPhone 11, Pro, Pro Max & MacBook: 2015+ & 1-8 \\
    \bottomrule
    \end{tabular}%
  \label{tab:participantsdetails}%
\end{table}%

 %\subsection{Appendix F: Recruitment on Facebook and LinkedIn text} 
% \label{recruitmentpost}
%Do you own and use an Apple Macbook and an iPhone? If yes, we would be interested in your experiences and perceptions on using both. We are conducting a research study at [institute and department details] on Investigating Cloud Platforms. To sign up for the study please click the link below for the pre-survey and we will contact you if you are eligible for this study. [Link to Survey].


\end{document}